\documentclass[preprint,12pt]{elsarticle}

\usepackage[utf8]{inputenc} 
\usepackage[T1]{fontenc}    
\usepackage{hyperref}       
\usepackage{url}            
\usepackage{booktabs}       
\usepackage{amsfonts}       
\usepackage{nicefrac}       
\usepackage{microtype}      
\usepackage{xcolor}         
\usepackage{placeins}  
\usepackage{amsmath}
\usepackage{array}
\usepackage{comment}
\usepackage{xcolor}

\usepackage{caption}
\usepackage{multirow} 
\usepackage{algpseudocode}
\usepackage{algorithm}
\usepackage{graphicx}
\usepackage{varwidth}%
\usepackage{hyperref}
\usepackage{xurl}
\biboptions{sort&compress}

\usepackage{xcolor}

\journal{Nuclear Physics B}

\begin{document}

\begin{frontmatter}



\title{NeuralFVM: Neural-physics-based Finite Volume Method for Turbulent Flows Using the $k$-$\omega$ Model}


\author[inst1]{Tingkai Xue}
\author[inst1]{Yu Jiao}
\author[inst2]{Te Ba}
\author [inst1] {Jingliang Wang}
\author[inst3]{Juntao Yang}
\author[inst3]{Simon See}
\author[inst4]{Boyang Chen}
\author[inst4]{Claire E. Heaney}
\author[inst4]{Christopher C. Pain}
\author[inst2]{Chang Wei Kang}
\author[inst1]{Mohamed Arif Bin Mohamed}
\author {Hongying Li$^{a*}$}

\affiliation[inst1]{organization={School of Mechanical and Aerospace Engineering, Nanyang Technological University},
addressline={50 Nanyang Avenue},
city={Singapore},
postcode={639798},
country={Republic of Singapore}}

\affiliation[inst2]{organization={Institute of High Performance Computing (IHPC), Agency for Science, Technology and Research (A*STAR)},
addressline={1 Fusionopolis Way, \#16-16, Connexis},
city={Singapore},
postcode={138632},
country={Republic of Singapore}}

\affiliation[inst3]{organization={NVIDIA AI Technology Centre, NVIDIA Corporation},
city={Singapore},
country={Republic of Singapore}}

\affiliation[inst4]{organization={Applied Modelling and Computation Group, Department of Earth Science and Engineering, Imperial College London},
addressline={South Kensington Campus, Exhibition Road},
city={London},
postcode={SW7 2AZ},
country={United Kingdom}}

\begin{abstract}
In this work, we develop a neural-physics solver based on finite volume method (FVM), namely NeuralFVM, for turbulent flows by implementing the standard $k$-$\omega$ model designed for efficient Graphics Processing Unit (GPU) execution. The governing equations for fluid flow and heat transfer are reformulated as local tensor operations using convolution-based stencil operators, which enables compatibility with deep learning libraries while preserving the conservative properties of the FVM. A key challenge in implementing the turbulent model within such a framework is the treatment of the stiff destruction terms in the $k$ and $\omega$ transport equations. To address this issue, an operator-splitting strategy is introduced in which the stiff destruction terms are handled semi-implicitly while the remaining terms are advanced explicitly. This formulation avoids global matrix assembly and allows the entire solver to be implemented using local tensor operations. In addition, the pressure-velocity coupling is solved using a convolution-based geometric multigrid algorithm embedded within a neural network architecture. The resulting NeuralFVM solver is validated through comparison with simulations conducted using the commercial CFD software ANSYS Fluent for several channel-flow configurations and an indoor airflow scenario. The results demonstrate close agreement in velocity, temperature, and turbulence quantities, confirming the accuracy of the proposed approach. The developed GPU framework achieves a speedup of around 19-46 times compared with its Central Processing Unit (CPU) counterpart under different meshes. Moreover, the proposed solver naturally integrates with machine learning workflows, providing a promising foundation for future data-driven turbulence modeling and optimization.
\end{abstract}


\end{frontmatter}


\section{Introduction}
Computational fluid dynamics (CFD) is widely used to analyze fluid flow and heat transfer by numerically solving the governing equations of fluid momentum and energy transport. Despite its widespread use, CFD simulations can be time-consuming and computationally expensive, particularly when turbulent flows are involved \cite{dns_tool}. As a result, considerable research efforts have focused on developing approaches that enable faster and more efficient CFD computations. Part of this improvement can be attributed to advances in hardware, in particular the development of Graphics Processing Units (GPUs), which enables faster and highly parallel matrix operations \cite{Ian_2024} compared to traditional Central Processing Unit (CPU) hardware. For example, significant acceleration has been reported when running ANSYS Fluent on NVIDIA GPUs \cite{Ian_2024}. Nevertheless, such hardware-based acceleration alone does not fully address the computational challenges of CFD. In addition, current implementations remain limited in versatility and, more importantly, further developments are needed to better support the integration of modern machine learning (ML) methodologies within CFD frameworks \cite{turbulence_age_of_data,ml_for_fluid_mechanics,CARON2025112229}.

Recent developments in automatic differentiation (AD) and GPU-based scientific computing have led to the emergence of differentiable and neural-physics solvers \cite{NP4Buoyancy}. High-level libraries provide interfaces that avoid low-level GPU programming while enabling gradient-based analysis. Researchers have leveraged these tools to construct CFD solvers compatible with differentiable programming frameworks \cite{Kochkov2021-ML-CFD,jaxfluid1,holl2024phiflow}. In particular, Chen et al. \cite{neural_physics} demonstrated that spatial discretization operators can be represented using convolutional neural network (CNN) layers, forming a neural-physics framework for solving the incompressible Navier-Stokes equations. Rather than learning parameters from data, the convolutional kernels are designed to directly represent the discretized coefficients of the governing equations. This approach preserves the underlying physics while reducing computational cost and allowing integration with ML workflows. Such neural-physics approaches have been successfully applied to a range of problems, including multiphase flows \cite{neural_multiphase}, reactor physics \cite{neural_neutron}, and particle packing \cite{neural_dem}.

Neural-physics first focused on solving the Navier-Stokes equations directly \cite{neural_physics}. For turbulent flows, this approach amounts to performing Direct Numerical Simulation (DNS). However, DNS is rarely used in engineering practice because of its prohibitive computational cost. An implicit large eddy simulation (LES) model has been implemented within neural-physics and applied to flow around a train \cite{Chen2025train}, along with several studies also implementing LES within differentiable, GPU-based CFD frameworks using JAX, applied to compressible turbulent flow \cite{jaxfluid2}. Although LES reduces computational overheads compared to DNS, it still requires fine spatial resolution and small time steps. Consequently, most industrial CFD simulations rely on Reynolds-averaged Navier-Stokes (RANS) models \cite{TOMINAGA2024105741}, which provide a more practical compromise between computational efficiency and accuracy. Most RANS models close the governing equations using the Boussinesq eddy viscosity hypothesis. This relates the Reynolds stress tensor to the mean strain-rate tensor through an eddy viscosity. Common examples include the $k$-$\varepsilon$\cite{LAUNDER1974269}  and $k$-$\omega$ \cite{Wilcox_2010} models. It introduces transport equations for the turbulent kinetic energy and dissipation rate or specific dissipation rate, and the Spalart-Allmaras model \cite{spalart1992one}, which was developed for aerodynamic flows. Existing efforts to incorporate this class of turbulence models into differentiable frameworks have primarily relied on data-driven approaches  \cite{Ling_Kurzawski_Templeton_2016, doi:10.2514/6.2017-1712, AGRAWAL2024112982}. 

Simulating RANS models requires solving additional transport equations for turbulent quantities. These equations usually contain stiff destruction terms that can produce negative values of turbulence variables in numerical discretizations, leading to numerical instability and divergence. To overcome this problem, methods such as the unconditionally positive-convergent implicit scheme can be used  \cite{MORYOSSEF20091984}. However, implicit schemes typically involve matrix inversion, which are difficult to implement within the neural-physics framework, where computational operations are required to remain local (as discussed in Section \ref{method:notation_tensor}). A similar problem is investigated by Zhao et al.\cite{ZHAO1997487}, who linearized the destruction terms to obtain a partial implicit method to perform time marching for the $k$-$\varepsilon$ model. Nevertheless, this approach requires a sufficiently small time step to ensure physically realistic values for the turbulence quantities.   

In this work, we develop a neural-physics solver, NeuralFVM, based on finite volume method (FVM) that implements the standard $k$-$\omega$ turbulence model on GPU architecture. The classical geometric multigrid (GMG) algorithm is reformulated using fixed-weight convolutional operators. The convolution kernels, including those used for smoothing and inter-grid transfer, are derived analytically from the governing equations rather than learned from data. The GMG procedure is implemented using an approach analogous to the U-Net architecture \cite{unet} commonly used for image segmentation tasks. To address the stiff destruction terms in the $k$-$\omega$ turbulence transport equations, an operator-splitting approach is proposed.

The novelty of the present study lies in four main aspects. First, we develop a RANS solver specifically formulated for efficient execution on GPUs, enabling substantial computational acceleration compared with conventional CPU-based CFD solvers. Second, the turbulence closure is implemented entirely through local tensor operations, eliminating the need for global matrix assembly and allowing the solver to fully exploit GPU parallelism. Third, the proposed NeuralFVM framework provides a fully differentiable, GPU-native RANS solver based on local tensor operations, enabling seamless integration with modern ML workflows and offering an efficient platform for data-driven modeling and optimization. Finally, the developed NeuralFVM can be executed on both GPU and CPU platforms with minimal modifications to the code, making it flexible and accessible to a wide range of users.

Unlike traditional CFD solvers, the NeuralFVM framework is fully differentiable, allowing gradients of flow quantities with respect to model parameters or boundary conditions to be computed automatically. This capability opens opportunities for gradient-based design optimization and data-driven turbulence modeling. The key enabling component is the formulation of the time-advancement procedure. Instead of constructing and solving a global matrix system arising from the spatial and temporal discretization of the governing equations, the proposed method employs an operator-splitting strategy in which individual terms are evaluated sequentially. This formulation enables stable implicit treatment of the stiff destruction terms while retaining explicit updates for the remaining less stiff contributions. Consequently, the turbulence model can be implemented entirely using local tensor operations commonly available in deep learning and numerical computing libraries, facilitating GPU acceleration and future integration with ML approaches.

In the NeuralFVM solver, heat transfer is incorporated by solving the energy conservation equation to obtain the temperature field. The ability to simulate thermal transport is important in many engineering applications, including occupant thermal comfort in buildings \cite{CATALINA20091740} and vehicles \cite{ZHAO2026130489}, thermal management of data centers \cite{SZELIGA2026128982}, and optimization of heat fin designs \cite{MOUSTAFA2025127735}. Since the energy equation contains convection and diffusion terms similar to those in the momentum equations, it can be solved using the same numerical approach adopted for the flow variables.

This paper is structured as follows. The governing equations are introduced in Section \ref{method:kw}, from which, several important aspects are identified and discussed sperately using notations introduced in Section \ref{method:notation_tensor}. These aspects include the finite volume discretization of the convection and diffusion terms explained in Section \ref{method:fvm}, the algorithm used by the GMG pressure solver to satisfy the continuity equation in Section \ref{method:pressure_correction}, and the semi-implicit treatment of the stiff destruction terms in Section \ref{method:kw_stiff}. The implementation of our model is verified against the results from commercial software ANSYS Fluent 2025 R2 as well as experimental data, and the results are presented in Section \ref{results}. Finally, we conclude with a summary of the main findings in Section 4.

\section{Methodology}
\subsection{Governing equations}\label{method:kw}
In this study, the standard $k$-$\omega$ turbulence model is adopted to represent turbulent effects in the flow field. The standard $k$-$\omega$ model \cite{Wilcox_2010} is chosen because it performs well for wall-bounded flows and is capable of resolving near-wall turbulence without relying on wall functions, in contrast to other RANS models such as the $k$-$\varepsilon$ model \cite{LAUNDER1974269}. This facilitates a consistent implementation within the present NeuralFVM framework based on local tensor operations. Its governing equations include the continuity (Eq. \eqref{eq:ko_cont}), momentum (Eq. \eqref{eq:ko_mom}), turbulence transport (Eq. \eqref{eq:ko_k},\eqref{eq:ko_o}), and energy  (Eq. \eqref{eq:ko_temp}) equations:
\begin{align}
    \frac{\partial u_i}{\partial x_i}&=0,\label{eq:ko_cont}\\
    \frac{\partial u_i}{\partial t}&=\frac{\partial}{\partial x_j} \left(-u_i u_j - \frac{p}{\rho}\delta_{ij}+(\nu+\nu_t) \frac{\partial u_i}{\partial x_j}-\frac{2}{3}k \delta_{ij}\right),\label{eq:ko_mom}\\
    \frac{\partial k}{\partial t}&=\frac{\partial }{\partial x_j}\left(-ku_j+\left(\nu+\frac{\nu_t}{\sigma_k}\right)\frac{\partial k}{\partial x_j}\right)+G_k-Y_k,\label{eq:ko_k}\\
   \frac{\partial \omega}{\partial t}&=\frac{\partial }{\partial x_j}\left(-\omega u_j+\left(\nu+\frac{\nu_t}{\sigma_\omega}\right)\frac{\partial \omega}{\partial x_j}\right)+G_\omega-Y_\omega,\label{eq:ko_o}\\
    \frac{\partial T}{\partial t}&=\frac{\partial}{\partial x_j} \left(-Tu_j+\left(\alpha+\frac{\nu_t}{Pr_t}\right) \frac{\partial T}{\partial x_j}\right),\label{eq:ko_temp}
 \end{align}
where $u_i$ and $p$ are the Reynolds-averaged velocity components and pressure respectively, $\nu_t$ is the turbulent viscosity, $\rho$ is density, $t$ is time, $\sigma_k$ and $\sigma_\omega$ are turbulent Prandtl number for $k$ and $\omega$, which are the turbulent kinetic energy and specific dissipation rate respectively, $G_k$ and $G_\omega$ are the generation terms of $k$ and $\omega$, respectively, $Y_k$ and $Y_\omega$ are destruction terms of $k$ and $\omega$, respectively, $\delta_{ij}$ is the Kronecker delta, $T$ is temperature, $\alpha$ is the diffusivity coefficient, and $Pr_t$ is the turbulent Prandtl number and it is chosen as 0.85 \cite{10.1115/1.2911398} in the present work. 

The Reynolds stress tensor in the RANS equations is modelled using the Boussinesq eddy viscosity hypothesis, which relates the Reynolds stresses to the mean strain-rate tensor through the turbulent viscosity $\nu_t$. $\nu_t$ can be computed from Eq. \eqref{eq:ko_turb_diff}. A damping factor $\alpha^*$ (Eq. \eqref{eq:a_star}) is included to account for the low-Reynolds number correction,
\begin{equation}
    \nu_t=\alpha^* \frac{k}{\omega} \label{eq:ko_turb_diff},
\end{equation}

These formulations are based on the ANSYS Fluent's theory guide \cite{AnsysFluentTheory2025} since the present model is verified through direct comparison with ANSYS Fluent results. The implementation closely follows the standard $k$-$\omega$ turbulence with the low-Reynolds number correction option applied. Accordingly, the following Eq. \eqref{eq:a_star} is used to compute $\alpha^*$, which damps the turbulent diffusivity values near the walls,
\begin{align}
    \alpha^*&=\alpha^*_\infty \left(\frac{\alpha_0^*+Re_t/R_k}{1+Re_t/R_k}\right) ,\label{eq:a_star}\\
    Re_t&=\frac{k }{\nu\omega},\label{eq:Re_t}
\end{align}
where $Re_t$ is the turbulent Reynolds number and the model constants are $\alpha_\infty^*=1, \alpha_0^*=\frac{\beta_i}{3}$\,and $R_k=6$ \cite{AnsysFluentTheory2025}. 

The generation of turbulent kinetic energy $G_k$ and specific dissipation $G_\omega$ are computed by  
\begin{align}
    G_k&=\nu_t S^2, \label{eq:k_gen}\\
    G_\omega&=\alpha \frac{\omega}{k}G_k ,\label{eq:o_gen}\\
    \alpha &= \frac{\alpha_\infty}{\alpha^*} \left(\frac{\alpha_0+Re_t/R_\omega}{1+Re_t/R_\omega}\right),\label{eq:a}
\end{align}
where $\alpha_\infty=0.52, \alpha_0=\frac{1}{9},R_\omega=2.95$ \cite{AnsysFluentTheory2025}, $S$ is the magnitude of the strain rate tensor and it is given by 
\begin{align}
    S=\sqrt{2S_{ij}S_{ij}}. \label{eq:strain_rate_magnitude}
\end{align}
Here $S_{ij}$ is the strain rate and it is calculated by
\begin{align}
    S_{ij}=\frac{1}{2}\left(\frac{\partial u_i}{\partial x_j}+\frac{\partial u_j}{\partial x_i}\right).\label{eq:strain_rate}\
\end{align}

The destruction of turbulent kinetic energy $Y_{k}$ and specific dissipation $Y_\omega$ are given by
\begin{align}
    Y_k&=\beta^* f_{\beta^*}k\omega, \label{eq:k_des}\\
    f_{\beta^*}&=\begin{cases}
        1, &\chi_k\leq 0\\
        \frac{1+680\chi_k^2}{1+400\chi_k^2}, &\chi_k>0
    \end{cases} \quad \text{where } \chi_k=\frac{1}{\omega^3}\frac{\partial k}{\partial x_j}\frac{\partial \omega}{\partial x_j},\\
    \beta^*&=\beta_\infty^* \frac{4/15+(Re_t/R_\beta)^4}{1+(Re_t/R_\beta)^4},\label{eq:beta_star}\\
    Y_\omega &= \beta f_\beta \omega^2, \label{eq:o_des}\\
    f_\beta &= \frac{1+70\chi_\omega}{1+80\chi_\omega} \qquad \text{ where } \chi_\omega = \left\vert\frac{\Omega_{ij}\Omega_{jk}S_{ki}}{(\beta_{\infty}^* \omega)^3}\right\vert,
\end{align}
where $ \beta_\infty^*=0.09,\beta=\beta_i=0.072$ \cite{AnsysFluentTheory2025}, $\Omega_{ij}$ is the rotation rate tensors given by
\begin{align}
    \Omega_{ij}=\frac{1}{2}\left(\frac{\partial u_i}{\partial x_j}-\frac{\partial u_j}{\partial x_i}\right).\label{eq:rotation_rate}
\end{align}
The governing equations can be separated into transient, convection, diffusion, generation, and destruction terms. The following sections describe the numerical methods used to solve each term and their implementation within the NeuralFVM solver.

\subsection{Notations and tensor operations}\label{method:notation_tensor}
\begin{figure}
    \centering
  \includegraphics[width=1.0\linewidth]{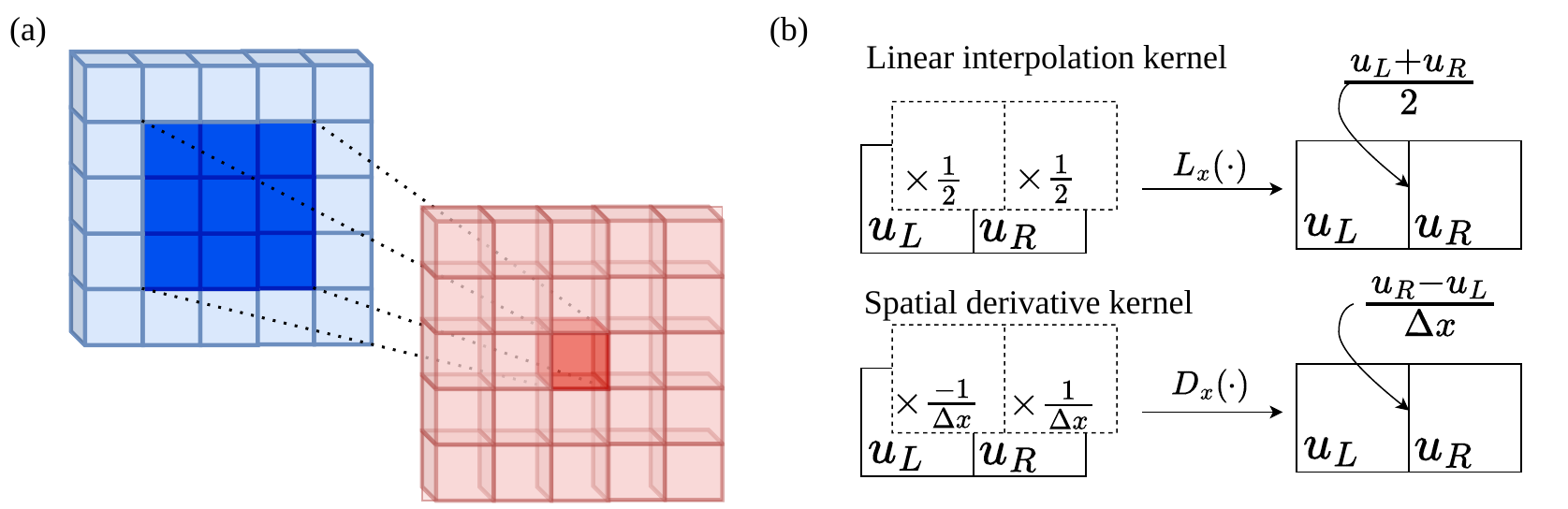}
    \caption{Illustrations of the neural-physics to be applied in CFD (a) convolutional kernels used for simulation of face-centered values from cell-centered values, (b) linear interploation and spatial derivative kernels.}
    \label{fig:kernel_examples}
\end{figure}

In this paper, we reformulate the finite-volume discretization of the governing equations as a sequence of local tensor operations, where the value computed at one point is only dependent on values at nearby points (illustrated in Fig. \ref{fig:kernel_examples}(a)). To aid in the discussion, the following notation is used for standardization.

Throughout this paper, physical variables follow the standard fluid mechanics notation. Specifically, the velocity components are represented by $u$, $v$, and $w$, pressure by $p$, temperature by $T$, turbulent kinetic energy by $k$, and the specific dissipation rate by $\omega$. Their derivatives with either time or space are denoted using subscripts (e.g., $\phi_t$ denotes the derivative with respect to time). All quantities will be implemented as tensors in PyTorch \cite{pytorch}. In alignment with the software implementation, these quantities should be viewed as a discretized field at a specific time (i.e. $u,v,w,p, T,k,\omega \in \mathbb{R}^{n_x\times n_y \times n_z}$). Each tensor element corresponds to the quantity value at a specific position. The tensors' shapes may differ depending on whether they include the boundary condition, and whether they correspond to face-centered or cell-centered values. To indicate that a tensor corresponds to face-centered rather than the cell-centered values, an additional subscript will be used as such $\phi\vert_{face}$.

Element-wise (Hadamard) product and division will be denoted by $\odot$ and $\oslash$ respectively. Multiplying all elements in $\phi$ by a scalar $k$ is denoted by $k\phi$. For addition and subtraction, the standard operators $+,-$ are used. In alignment with the syntax of the PyTorch library, the context is important in understanding the computation involved. For instance, consider the diffusion term in the momentum equation Eq. \eqref{eq:ko_mom}, the molecular viscosity $\nu$ is a scalar while the turbulent viscosity $\nu_t$ varies with space and therefore is stored as a tensor. Adding them would mean adding the same $\nu$ value to all elements of the $\nu_t$ tensor, while on the other hand, adding two tensors of the same size would mean an elementwise addition. As such, it is important to keep track of the meaning and shapes of the different tensors. 

We represent discrete numerical operators using convolutional kernels typically employed in CNNs. In this paper, convolutional kernels correspond to a compact stencil used to compute either spatial derivatives or linear interpolation at cell faces from cell centers or vice versa. This is in contrast to the finite-element formulation used by that of Chen et al.\cite{neural_physics} which solely computes nodal values. These operators are denoted using capital letters with a subscript that indicates the relevant dimension, while the application of these operators will be denoted as $\ast$. For example, the application of convolutional operators that compute derivative and perform linear interpolation along the $x$ direction will be denoted as $D_x$ and $L_x$ (Eq. \eqref{eq:Dx_Lx}) respectively:
\begin{align}
    x\in\mathbb{R}^{n_x\times n_y \times n_z}, D_x \ast x\in \mathbb{R}^{(n_x-1)\times n_y \times n_z}, L_x\ast x \in \mathbb{R}^{(n_x-1)\times n_y \times n_z} .\label{eq:Dx_Lx}
\end{align}
Note that tensor shapes change after these operations. This is consistent with the behaviour of CNN layers when no padding is applied to restore the original tensor dimensions. From a physical standpoint, these operations compute the face-centered values from cell-centered values, which reduces the tensor size along the corresponding dimension by one. An illustration is provided in Fig. \ref{fig:kernel_examples}(b);

\begin{figure}
    \centering
    \includegraphics[width=1.0\linewidth]{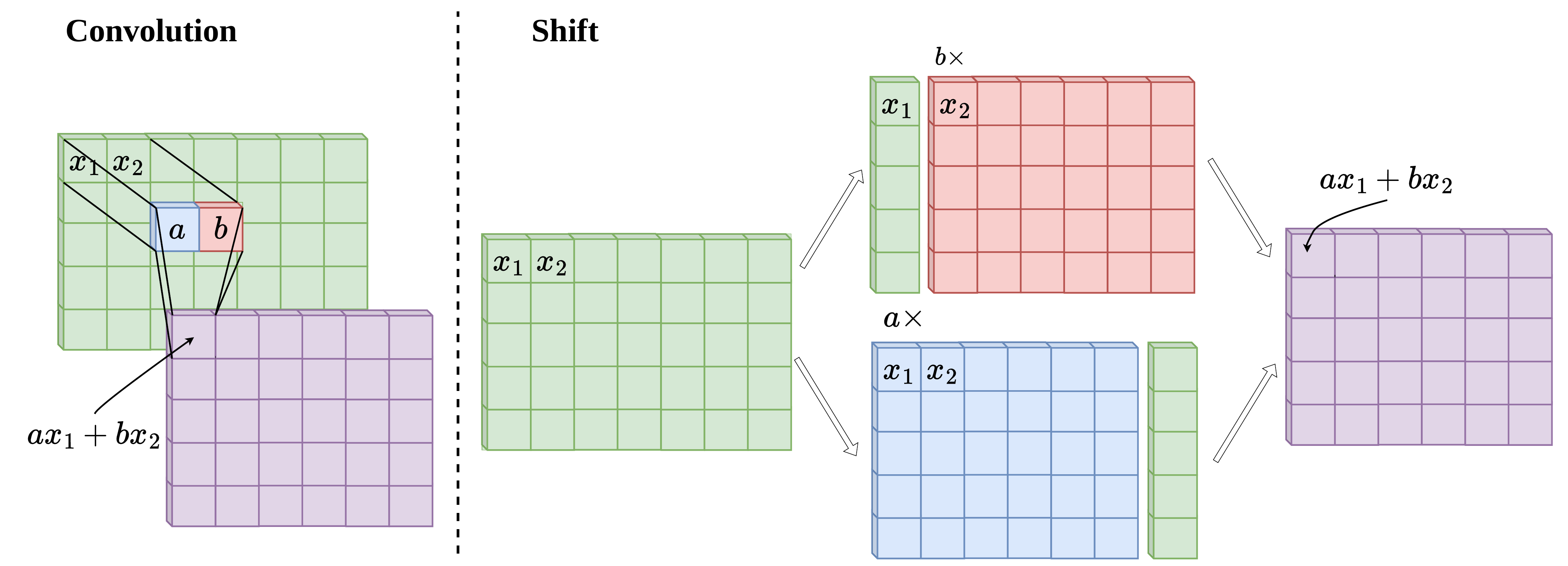}
    \caption{Illustration of convolutional operations implemented through simple addition or subtraction of shifted tensors.}
    \label{fig:convolution_v_shift}
\end{figure}

Since the kernels employed in this work have compact stencils, performing direct convolution operations is not computationally optimal. Instead, the equivalent stencil operations are implemented using tensor slicing and shifting, which yield identical numerical results while improving computational efficiency. This process is illustrated in Fig. \ref{fig:convolution_v_shift}. To obtain the purple tensor from the green tensor, the green tensor is first sliced to produce the red and blue tensors, which are then summed. For consistency in terminology, these stencil operations are still referred to as convolution throughout the paper. Hereafter, the proposed framework is referred to as NeuralFVM.

\subsection{FVM with upwind scheme}\label{method:fvm}
The FVM is used to discretize all the governing equations in Section \ref{method:kw}. The convection term is discretized using the first-order upwind scheme, which is widely adopted to ensure numerical stability in strongly convective flows \cite{cfd_fvm}. In the implementation, the upwind and downwind operators are defined to select the value in the adjacent grid cell corresponding to a lower and higher $x$ coordinate respectively. These operations can also be implemented by defining convolutional kernels with values $[1,0]$ and $[0,1]$. Notation wise, the convolutional operators that take the upwind and downwind values along the $x$ direction are denoted by $W_x^+$ and $W_x^-$ respectively. The selection of the cell value to take can be controlled by an indicator variable $I(\cdot)$ that checks whether the face-centered velocity value is positive or negative. As such, the upwind convective flux along the $x$-direction is computed as
\begin{align}
    \phi_{face,x}^{upwind} &\gets (W_x^+\ast BC_x(\phi)) \odot I[u_{face,x}\geq 0] \nonumber\\
    &\qquad+(W_x^-\ast BC_x(\phi))\odot I[u_{face,x} < 0] ,\nonumber\\
    (u\phi)^{upwind}_{face,x}&\gets \phi^{upwind}_{face,x}\odot u_{face,x} ,\label{eq:upwind_x}
\end{align}
where $BC_x$ is an operation that applies the boundary condition along the $x$-direction (explained in Section \ref{method:boundary_conditions}). The values for the other directions can be obtained in a similar manner. The algorithm for extracting the upwind convective term is described in Alg. \ref{algo:upwind}. It is necessary to mention the shapes of the tensors are consistent. The diffusive flux is computed using the central difference scheme. After computing the combined face-centered flux values $f_x,f_y,f_z$, integration over the control volume is performed according to the divergence theorem: 

\begin{align}
    \frac{\partial \phi}{\partial t}\gets D_x\ast f_x+D_y\ast f_y + D_z\ast f_z .\label{eq:divergence_theorem}
\end{align}

\subsection{Multigrid pressure and velocity correction} \label{method:pressure_correction}
The coupling between velocity and pressure is treated using a projection method \cite{projection}. After advancing the velocity field $\mathbf{u}=[u,v,w]$ using the time integration scheme explained in Section \ref{method:time_integration}, we obtain an intermediate velocity field $\mathbf{u}^*$. This intermediate velocity $\mathbf{u}^*=[u^*,v^*,w^*]$ may not satisfy the divergence-free condition, giving rise to a cell-centered residual field $r$ obtained from discretizing the rate of change of the divergence of the velocity field:
\begin{align}
    -\frac{\partial \nabla \cdot \mathbf{u}^*}{\partial t}&\approx-\frac{1}{\Delta t}\left(D_x\ast BC_x(u^*)+D_y\ast BC_y(v^*) +D_z\ast BC_z(w^*)\right)\\
    &=r, \label{eq:pc_residual}
\end{align}
where $BC_x,BC_y,BC_z$ corresponds to functions that apply the boundary condition on the velocity field via padding (explained in Section \ref{method:boundary_conditions}).

To obtain a divergence-free velocity field, the pressure field is updated by solving the Poisson equation (Eq. \eqref{eq:pc_differential}), which when discretized gives 
\begin{align}
    -\frac{\partial \nabla \cdot \mathbf{u}^*}{\partial t}&=-\nabla^2 p \label{eq:pc_differential},\\
    r&= Ap ,\label{eq:pc_discretized}
\end{align}
where $A$ is a linear operator denoting the negative Jacobian operator. The value of the Jacobian can be obtained using the central-difference scheme for second derivatives, giving 

\begin{equation}
\begin{aligned}
(\nabla^2 p)_{i,j,k} \approx
& \frac{p_{i+1,j,k}-2p_{i,j,k}+p_{i-1,j,k}}{\Delta x^2} \\
& + \frac{p_{i,j+1,k}-2p_{i,j,k}+p_{i,j-1,k}}{\Delta y^2} \\
& + \frac{p_{i,j,k+1}-2p_{i,j,k}+p_{i,j,k-1}}{\Delta z^2},
\end{aligned}
\label{eq:neg_lap_discretize}
\end{equation}
where the subscripts denote the cell indices along the $x$, $y$, and $z$ directions. Notably, the value at a given cell is computed solely from the values of its neighboring cells. This locality implies that the computation can be naturally implemented as a convolution operation. In the following, $A$ is denoted either the matrix operator or the convolution operator, depending on whether the convolution symbol 
$*$ is used. The discretized operation can therefore be rewritten as
\begin{align}
    r=A\ast BC(p),
\end{align}
where $BC$ denotes the function that apply boundary conditions on all three directions via padding i.e. $BC(\cdot)=BC_x(BC_y(BC_z(\cdot )))$. The values of the 3D kernel when $\Delta x=\Delta y=\Delta z$ are 
\begin{align}
    -\frac{1}{\Delta x^2}\left[\:\begin{bmatrix}
        0&0&0\\
        0&1&0\\
        0&0&0
    \end{bmatrix}, \begin{bmatrix}
        0&1&0\\
        1&-6&1\\
        0&1&0
    \end{bmatrix}, 
    \begin{bmatrix}
        0&0&0\\
        0&1&0\\
        0&0&0
    \end{bmatrix}\:\right].
\end{align}

Eq. \eqref{eq:pc_discretized} is solved iteratively using the the Jacobi method, which is described by the iterative equation 
\begin{align}
    p^{(k+1)}=p ^{(k)}-D^{-1}Ap^{(k)}+D^{-1}r^{(k)}, \label{eq:jacobi_pc}
\end{align}
where $D$ is a diagonal matrix whose entries correspond to the diagonal elements of $A$. Due to the structure of $A$, $D$ can be replaced with a scalar of value $\frac{6}{\Delta x^2}$. 

\begin{figure}
    \centering
    \includegraphics[width=1.0\linewidth]{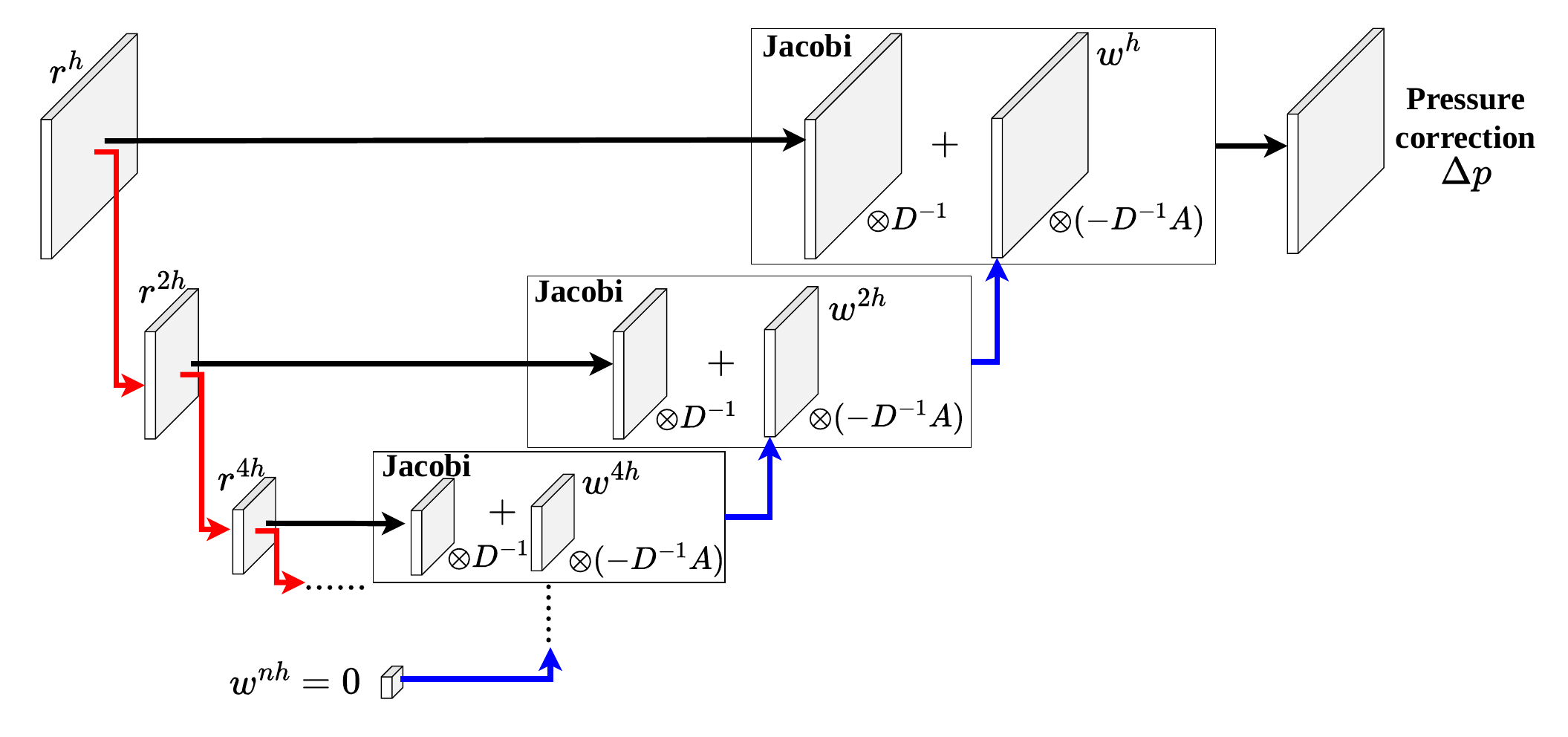}
    \caption{Pressure correction using the Jacobi method coupled with a multigrid approach. Red and blue arrows indicate restriction and prolongation respectively. $\otimes$ indicate the left multiplication of a matrix, which can be implemented using convolutional layer.}
    \label{fig:pc_multigrid}
\end{figure}

 However, for large-scale CFD problems, the pressure Poisson equation can be computationally expensive to solve. To improve efficiency, a multigrid strategy is adopted, leveraging the structural similarity between classical multigrid solvers and the U-net architecture \cite{unet}. In the multigrid approach, the residual on finer mesh is converted to the one on coarser mesh (restriction), where iterative updates are performed, and converted back to the finer mesh (prolongation). This structure is similar to the U-net architecture, where deeper levels contain lower resolution, but larger scale information (Fig. \ref{fig:pc_multigrid}). Notation wise, restriction (RES) and prolongation (PROL) operators are written as

\begin{align}
    RES&: \mathbb{R}^{n_x\times n_y\times n_z} \rightarrow \mathbb{R}^{\frac{n_x}{2}\times \frac{n_y}{2}\times \frac{n_z}{2}},\label{eq:restriction}\\
    PROL&:\mathbb{R}^{\frac{n_x}{2}\times \frac{n_y}{2}\times \frac{n_z}{2}}\rightarrow \mathbb{R}^{n_x\times n_y\times n_z},\label{eq:prolongation}
\end{align}
where RES performs grid coarsening and PROL interpolates the coarse-grid correction back to the fine grid. These operations can be implemented using pooling and upsampling layers available in standard deep learning libraries. The algorithm is provided in Alg. \ref{algo:pressure_correction}.
\begin{algorithm}[h]
  \caption{Multigrid pressure correction}\label{algo:pressure_correction}
  \begin{algorithmic}[1]
    \Function{MGPressureCorrection}{$u,v,w,p,\Delta t, n_{iter},n_{level}$}
        \State $\bar{u}\gets BC_x(u);\bar{v}\gets BC_y(v);\bar{w}\gets BC_z(w)$
        \State $u\vert_{face,x}\gets L_x\ast \bar{u}$
        \State $v\vert_{face,y}\gets L_y \ast \bar{v}$
        \State $ w\vert_{face,z}\gets L_z\ast \bar{w}$
       \State $b=-\frac{D_x\ast u\vert_{face,x}+D_y\ast v\vert_{face,y}+D_z\ast w\vert_{face,z}}{\Delta t}$
       \State $r_0\gets A\ast BC(p)-b$
       \For{$i\in \{1,...,n_{iter}\}$}
            \State $w\gets 0$
            \For{$l\in \{1,...,n_{level}\}$}
                    \State $r_l \gets RES(r_{l-1})$
            \EndFor
            \For{$l\in \{n_{level},...,1\}$}
                    \State $w \gets w-D^{-1}A\ast BC(w)+D^{-1}r_l$
                    \State $w\gets PROL(w)$
            \EndFor
            \State $p\gets p-w$
            \State $p\gets p-D^{-1}A\ast BC(p)+D^{-1}b$
       \EndFor
       \State \Return $p$
    \EndFunction
  \end{algorithmic}
\end{algorithm}

\subsection{Computing the stiff destruction terms}\label{method:kw_stiff}
Simulating the $k$-$\omega$ turbulence model can be challenging due to the stiff destruction terms, justifying the usage of implicit methods for simulating turbulent flows in traditional numerical schemes \cite{nasa_sti}. However, these methods cannot be implemented directly using our neural-physics framework since implicit methods usually involve matrix inversions, which causes the operation to lose the local property (Fig. \ref{fig:kernel_examples}(a)). A similar problem is investigated by Zhao et al. \cite{ZHAO1997487}, where a partial implicit method is proposed to perform time marching for the $k$-$\varepsilon$ turbulence model. However, there is still a limit to the time step required to ensure realistic values of the turbulence quantities are obtained. In this work, we propose a different approach to overcome this limitation.

Since the stiffness of the equations mainly arises from the destruction terms for $k$ and $\omega$, it is natural to use splitting method \cite{splitting} to split the transport equations for $k$ (Eq. \eqref{eq:ko_k}) and $\omega$ (Eq. \eqref{eq:ko_o}) into the destruction terms (denoted with a subscript $des$) and the rest of the terms (consolidated into the term with subscript $rest$) and handle them separately:

\begin{align}
    &\frac{\partial k}{\partial t}=\left(\frac{\partial k}{\partial t}\right)_{des}+\left(\frac{\partial k}{\partial t}\right)_{rest},\\
    &\frac{\partial \omega}{\partial t}=\left(\frac{\partial \omega}{\partial t}\right)_{des}+\left(\frac{\partial \omega}{\partial t}\right)_{rest},\\
    &\left(\frac{\partial k}{\partial t}\right)_{des}=-Y_k=-\beta^* f_{\beta^*}k\omega, \\
    &\left(\frac{\partial \omega}{\partial t}\right)_{des}=-Y_\omega = -\beta f_{\beta} \omega^2.
\end{align}

As the destruction terms turbulent fields themselves,  the equations can be discretized using the semi-implicit Euler's method to give

\begin{align}
    \frac{k^{n}_1-k^n}{\Delta t}&=-\beta^* f_{\beta^*} k^{n}_1\omega ^n \Longrightarrow k_1^{n}=\frac{k^n}{1+\Delta t \beta^* f_{\beta^*}\omega^n} ,\label{eq:k_des_step}\\
    \frac{\omega_1^{n}-\omega^n}{\Delta t}&=-\beta f_{\beta} \omega_1^{n}\omega^n \Longrightarrow \omega_1^{n}=\frac{\omega^n}{1+\Delta t \beta f_{\beta} \omega^n}, \label{eq:o_des_step}
\end{align}
$k_1^n$ and $\omega_1^n$ denote the values of $k$ and $\omega$ after advancing the destruction terms by $\Delta t$. Our combined time integration scheme involves separately advancing for a time step of $\frac{1}{2}\Delta t$ twice. Eq. \eqref{eq:k_des_step} and \eqref{eq:o_des_step} can also be easily implemented using elementwise operations as
\begin{align}
    k_1^n &\gets k^n \oslash (1+\Delta t \beta^* f_{\beta^*}\omega^n), \label{eq:k_des_tensor}\\
    \omega_1^n &\gets \omega^n \oslash (1+\Delta t \beta f_{\beta}\omega^n). \label{eq:o_des_tensor}
\end{align}
It should be noted that this method of advancing the destruction terms is unconditionally stable, and the values remain positive, making the time step limited by the much more forgiving CFL condition imposed by the explicit scheme used to advance the rest of the terms. For all the cases presented below, the time step size of  $\Delta t=2\times 10^{-3}$ s is adopted and has been verified to ensure numerical stability, unless otherwise specified. 

\subsection{Time integration} \label{method:time_integration}
\begin{figure}
    \centering
    \includegraphics[width=0.9\linewidth]{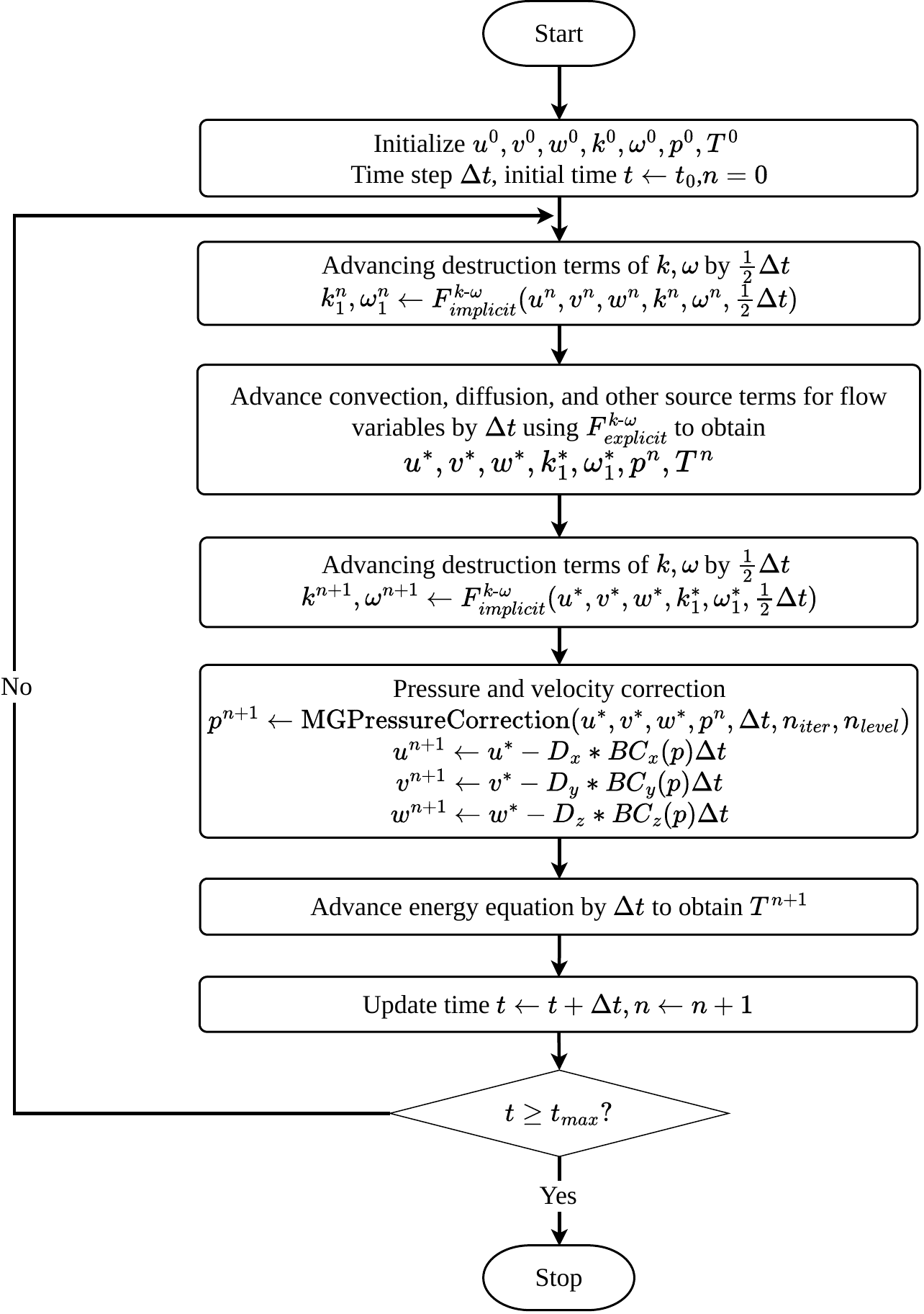}
    \caption{Flow chart showing the sequence of computation in one time advancement step}
    \label{fig:framework}
\end{figure}

Our NeuralFVM framework is a transient solver that performs time advancement via time marching. Within each time step, the stiff desctruction terms are treated semi-implicitly (as explained in Section \ref{method:kw_stiff}) while the rest of the terms are solved using standard explicit Runge-Kutta schemes. To combine the implicit and explicit steps, Strang splitting could be used, where an operator is split into a symmetric sequence \cite{splitting}. In our time integration scheme, one full time step of the explicit component is performed in between two half time steps of the semi-implicit component. A flow chart showing the computation involved in one time advancement step is illustrated in Fig. \ref{fig:framework}. 
\subsection{Boundary conditions}\label{method:boundary_conditions}
Boundary conditions (BCs) are enforced using ghost cells, which are virtual cells outside of the computational domain used to enforce boundary conditions. These boundaries can be either external or internal. Internal boundaries arise when solid blocks are embedded within the domain, forming interfaces between different regions. Consequently, the ghost cells may lie either outside or inside the spatial domain of the tensors.

\begin{figure}
    \centering
    \includegraphics[width=1.0\linewidth]{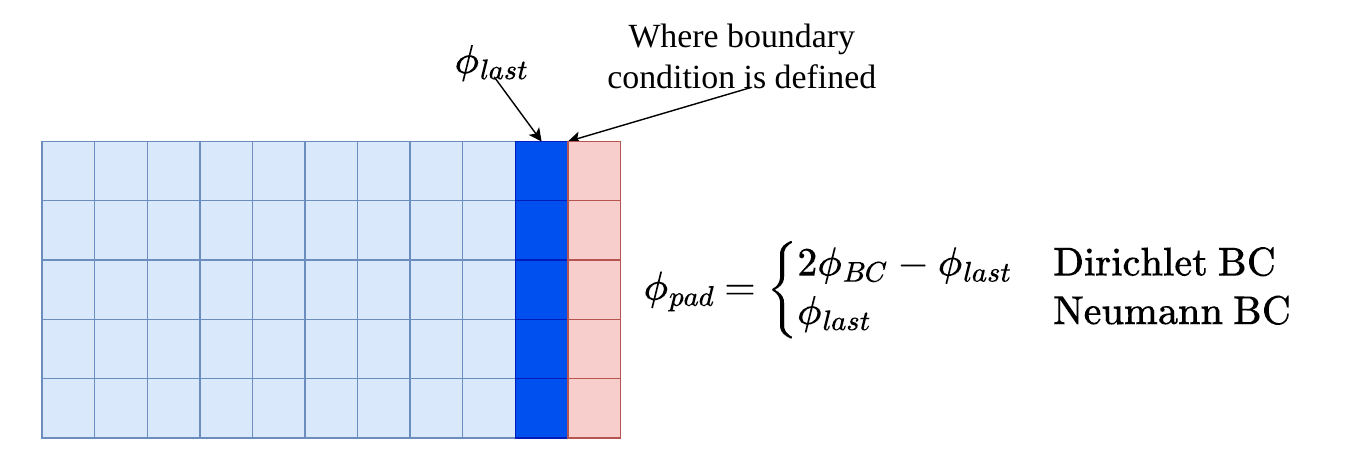}
    \caption{External BCs are implemented by padding the tensor with values that enforce the specified boundary conditions.}
    \label{fig:boundary_conditions}
\end{figure}

To impose external BCs, the computational tensors are padded with a single layer of ghost cells, whose values are prescribed according to the specified boundary condition. The implementation is illustrated in Fig. \ref{fig:boundary_conditions}. The operators for applying BCs along each of the three spatial directions are denoted as
\begin{align}
    &BC_x:\mathbb{R}^{n_x\times n_y\times n_z}\rightarrow \mathbb{R}^{(n_x+2)\times n_y\times n_z},\label{eq:bc_x}\\
    &BC_y:\mathbb{R}^{n_x\times n_y\times n_z}\rightarrow \mathbb{R}^{n_x\times (n_y+2)\times n_z},\label{eq:bc_y}\\
    &BC_z:\mathbb{R}^{n_x\times n_y\times n_z}\rightarrow \mathbb{R}^{n_x\times n_y\times (n_z+2)},\label{eq:bc_z}\\
    &BC:\mathbb{R}^{n_x\times n_y\times n_z}\rightarrow \mathbb{R}^{(n_x+2)\times (n_y+2)\times (n_z+2)}.\label{eq:bc}
\end{align}

\begin{figure}
    \centering
    \includegraphics[width=1.0\linewidth]{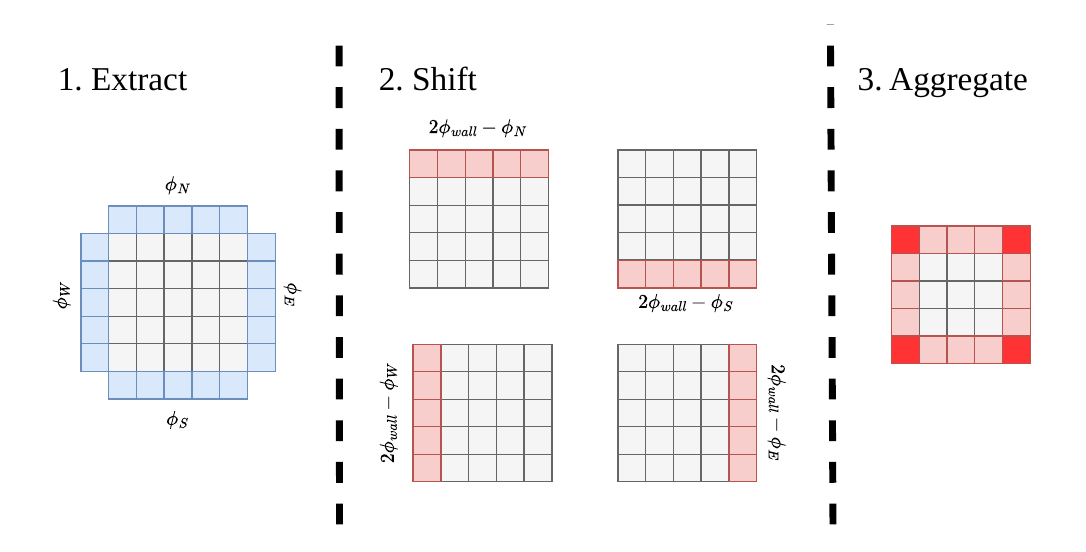}
    \caption{Internal BCs are imposed by extracting the neighboring fluid values, shifting them appropriately, and aggregating them to compute the required quantities near the interface.}
    \label{fig:internal_boundary_conditions}
\end{figure}

Internal BCs are required when solid blocks are present within the computational domain represented by the tensors. These conditions are also enforced using ghost cells. Unlike external boundaries, however, the location of the solid interface must be explicitly tracked. This is achieved by precomputing masks that identify the solid boundaries, then extracting the  relevant neighboring fluid values, shift them accordingly, and aggregating them to obtain the required quantities near the interface. A 2D example is shown in Fig. \ref{fig:internal_boundary_conditions}. 

The no-slip BC is applied at the walls. For $k$, ANSYS Fluent's treatment of  $\frac{\partial k}{\partial n}=0$ is used. In this study, the viscous sublayer is resolved ($y^+\sim 1$) and the correlation $\omega_{wall}=\frac{6\nu}{\beta_i d^2}$ is used, where $d$ is the distance between the cell centroid and the wall. It should be noted that this treatment differs from newer versions of Fluent, such as 2025 R2 used in this study, which employ calibrated $y+$ insensitive wall treatments. This difference may explain some discrepancies in the results. Nonetheless, the present formulation is retained for its simplicity. For temperature, a Dirichlet BC is applied at the walls.

\section{Results and Discussions}\label{results}
\subsection{Verifications}\label{results:verification}
To verify the implementation of the proposed algorithm, two channel-flow scenarios are investigated: an open-channel flow and a channel flow with blocks. The computational domain for both cases is $2.56\,\text{m} \times 0.64\,\text{m} \times 0.64\,\text{m}$. A uniform inlet velocity of $u_{inlet}=0.1m/s$ is prescribed at the inlet. The simulation is performed with kinematic viscosity $\nu=1.46\times 10^{-5}m^2/s$, corresponding to a Reynolds number of $Re=4384$. A fully developed outflow condition is imposed at the outlet while no-slip boundary conditions are applied on all walls. The temperature of the fluid at the inlet is $25^{\circ}C$ and the wall temperature is maintained at $35^{\circ}C$.

The same geometries and boundary conditions are implemented on ANSYS Fluent. Incompressible flow simulations are performed using the pressure-based solver. For spatial discretization, the Least Squares Cell Based method is used for gradient computation. The pressure equation is discretized using the second order scheme, while the momentum, turbulent kinetic energy ($k$), specific dissipation rate ($\omega$), and energy equations are all solved using the first order upwind scheme. 

A mesh independence study is performed using uniform mesh sizes of 1cm, 5 mm and 4 mm. The results indicate that 5mm and 4mm mesh sizes yield nearly identical velocity and temperature distributions, as well as similar turbulence quantities $k$ and $\omega$ profiles along the centerline. Consequently, the 5 mm mesh is used in the subsequent simulations. 
\begin{figure}
    \centering
    \includegraphics[width=1.0\linewidth]{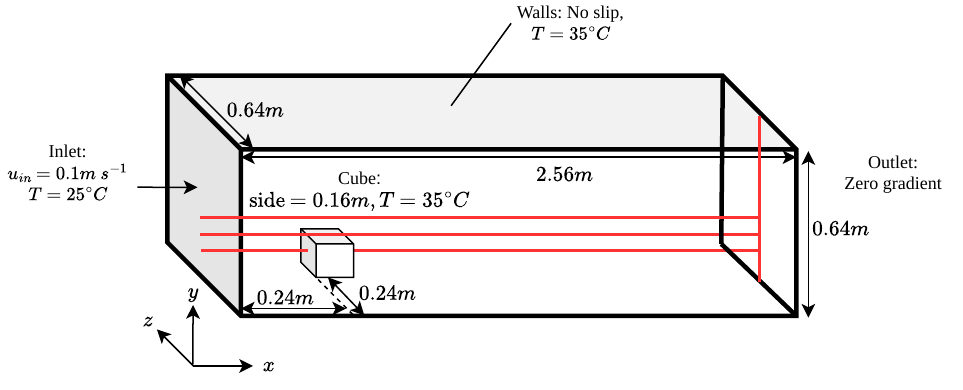}
    \caption{Channel flow scenario used for the verification of our NeuralFVM solver. The red lines indicate the locations where results are extracted for comparison with ANSYS Fluent.}
    \label{fig:verification_channel}
\end{figure}

It should be noted that the solution procedures differ between ANSYS Fluent and the proposed NeuralFVM solver. In ANSYS Fluent, the steady-state solution is obtained directly, whereas NeuralFVM performs a transient simulation until it becomes statistically stationary. It should be noted that time averaging of the unsteady RANS simulations may yield different results from the steady RANS simulation \cite{Durbin_Reif_2011}. However, we verified that the time derivative terms vanished and that a stationary solution is achieved by continuing the simulation for an additional 1000 time steps, during which no observable change in the extracted solution is detected. The same procedure is applied to all other cases unless otherwise stated. 

\begin{figure}
    \centering
    \includegraphics[trim={9cm 0cm 15cm 0cm}, width=0.9\linewidth]{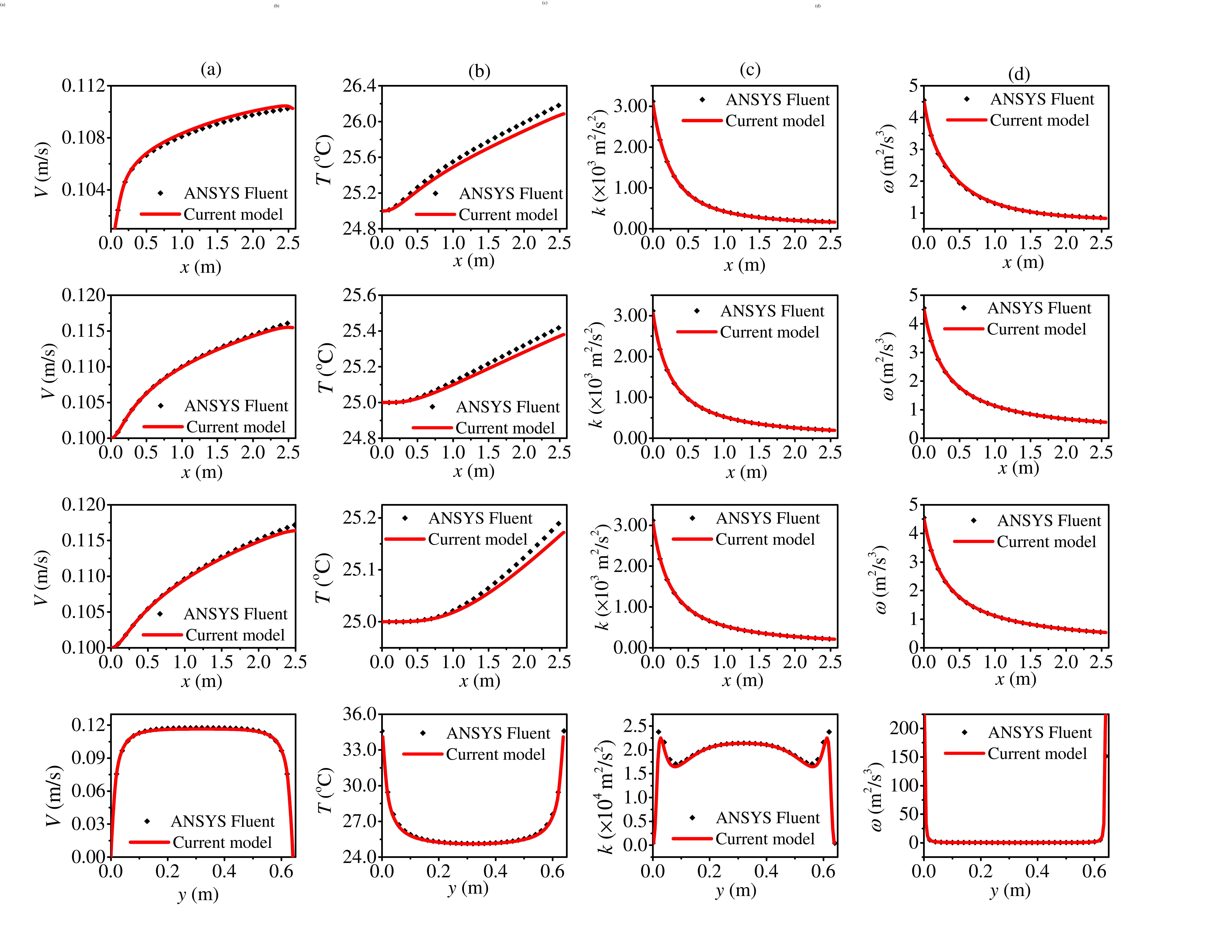}
    \caption{Comparison between the (a) velocity magnitude $V$, (b) temperature $T$, (c) turbulent kinetic energy $k$, and (d) specific dissipation rate $\omega$, between the NeuralFVM solver and the Fluent simulations for the open channel verification case. From top to bottom, the results are for the streamwise direction at $0.08 m$, $0.16 m$, and $0.24 m$ from the center of the bottom wall, and the center of the outlet.}
    \label{fig:results_openchannel}
\end{figure}

The velocity magnitude $V$ and temperature $T$ as well as $k$ and $\omega$ obtained from our simulations are evaluated along three streamwise lines located at $0.08 m$, $0.16 m$, and $0.24 m$ from the center of the bottom wall, as well as along a vertical line passing through the center of the outlet surface (illustrated as red lines in Fig. \ref{fig:verification_channel}). Fig.~\ref{fig:results_openchannel} shows a comparison between the results obtained from ANSYS Fluent and the proposed NeuralFVM solver. Good agreement is observed between the two approaches, even for the sharp gradients near the wall for both $k$ and $\omega$.

Further verification is performed for a channel with block as shown in Fig. \ref{fig:verification_channel}. Different from the open-channel flow, a cube with a side length of 0.16m is placed along the centerline of the channel at a distance of 0.24m from the inlet. This case is designed to assess the capability of the solver in handling internal walls within the computational domain. The boundary conditions remain the same as in the open-channel case. The cube surface is maintained at a temperature of $35^\circ C$. 

\begin{figure}
    \centering
    \includegraphics[trim={9cm 0 15cm 0}, width=0.9\linewidth]{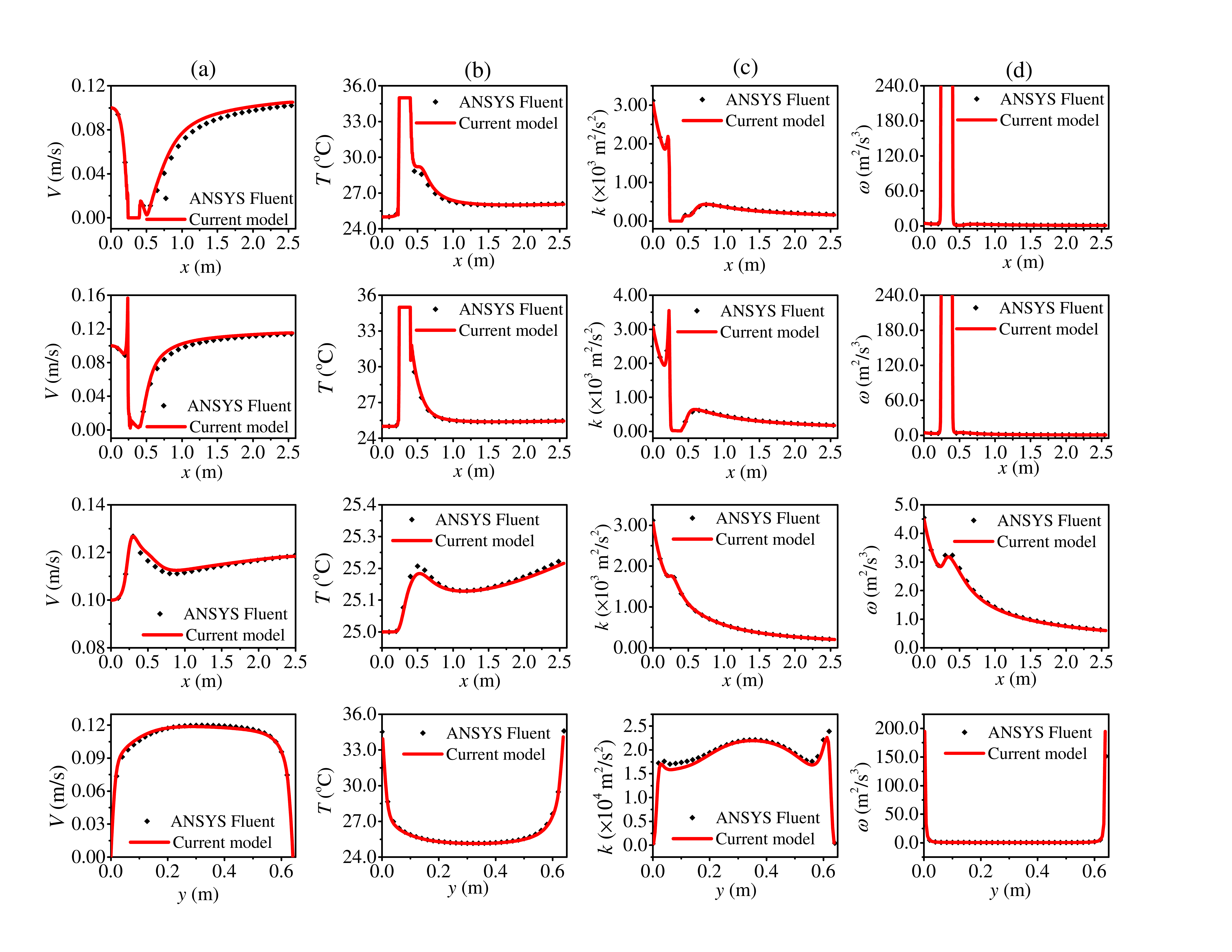}
     \caption{Comparison between the (a) velocity magnitude $V$, (b) temperature $T$, (c) turbulent kinetic energy $k$, and (d) specific dissipation rate $\omega$, between the NeuralFVM solver and the Fluent simulations for the blocked channel verification case. From top to bottom, the results are for the streamwise direction at $0.08 m$, $0.16 m$, and $0.24 m$ from the center of the bottom wall, and the center of the outlet.}
    \label{fig:results_blockchannel}
\end{figure}

Fig. \ref{fig:results_blockchannel} presents the comparison between the NeuralFVM predictions and the ANSYS Fluent solutions for the channel with a block case. The related quantities at the same locations as the open-channel case is chosen for comparsion purpose. The velocity and temperature distributions are well reproduced throughout the domain, including the rapid variations near the wall. The peak values and downstream decay of the turbulence quantities $k$ and $\omega$ are also well captured. Minor discrepancies are observed only in regions with steep gradients close to the wall, which may be attributed to differences in near-wall treatment and time integration methods. These results demonstrate that the proposed NeuralFVM solver can reproduce both mean flow and turbulence characteristics with accuracy comparable to a commercial CFD solver.     

\subsection{Case studies}
To further evaluate the capability of NeuralFVM under more challenging configurations, four additional cases are examined: (1) multiple aligned blocks of different sizes, (2) staggered blocks of different sizes, and (3) an array of blocks with identical size, and (4) the Annex 20 benchmark indoor airflow test.

\subsubsection {Multiple aligned blocks of different sizes}
\begin{figure}
    \centering
    \includegraphics[width=1.0\linewidth]{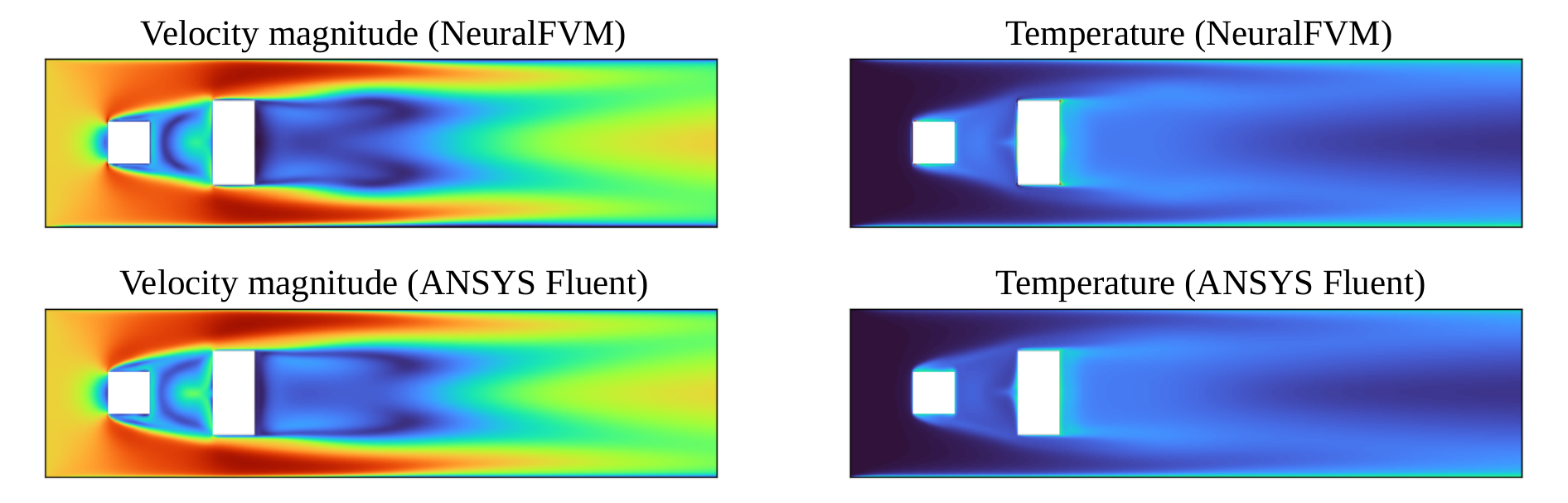}
    \caption{Comparison between the velocity magnitude $V$, temperature $T$, turbulent kinetic energy $k$, and specific dissipation $\omega$, between the NeuralFVM simulation and the Fluent simulations for the staggered blocks case}
    \label{fig:compare_two_blocks}
\end{figure}

In this case, two aligned blocks are placed along the channel centerline. The first block is a cube with a side length of 0.16m, followed downstream by a larger cuboid of dimensions $0.16\,\text{m} \times 0.24\,\text{m} \times 0.24\,\text{m}$. This configuration is designed to examine the solver\textquoteright s capability in handling multiple internal boundaries and the interaction of wakes generated by obstacles of different sizes. 
Fig. \ref{fig:compare_two_blocks} presents the velocity magnitude and temperature profiles obtained from NeuralFVM and ANSYS Fluent. The overall flow structures predicted by NeuralFVM are highly consistent with the ANSYS Fluent results. The NeuralFVM solver accurately captures the flow acceleration around the blocks, the separation and recirculation zones behind the two blocks, and the gradual recovery of the wake downstream. Since the block surfaces are maintained at a higher temperature than the incoming fluid, the surrounding fluid is heated and forms a downstream thermal wake along the channel centerline due to convective heat transport. This behavior is also clearly reproduced by NeuralFVM. It demonstrates the accuracy of the NeuralFVM solver.

\subsubsection {Staggered blocks of different sizes}
The second demonstration case focuses on two offset blocks within the computational domain. The domain dimensions are the same as those used in the previous case. The first block is a smaller cube (with sides $0.16m$) located closer to the left sidewall, while the second block is a taller cuboid (with height $0.32m$) positioned slightly downstream and laterally offset from the first. The streamwise distance between the front of the first and second block is $0.26m$, and the lateral offset distance between the sides is $0.08m$. This configuration is selected to evaluate the model\textquoteright s capability to accurately resolve multiple internal solid boundaries, as well as to capture the complex flow interaction and wake development induced by obstacles of different heights within a confined distance.

\begin{figure}
    \centering
    \includegraphics[width=1.0\linewidth]{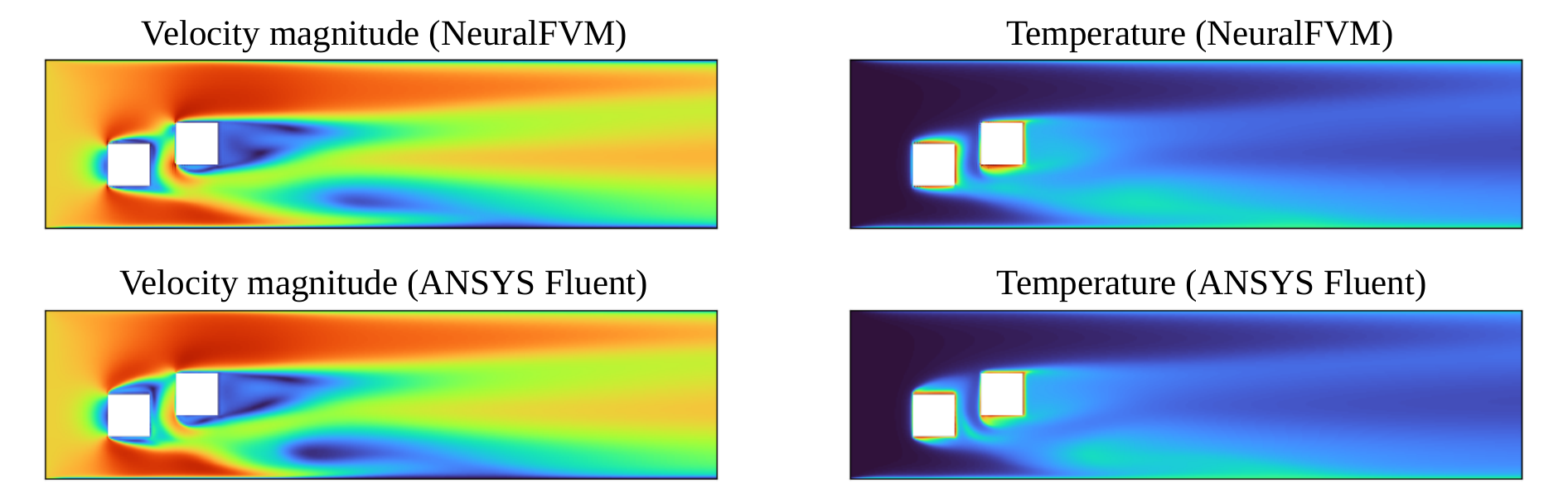}
    \caption{Comparison of velocity magnitude and temperature profiles of simulation of the staggered blocks case between NeuralFVM and Fluent}
    \label{fig:compare_staggered_blocks}
\end{figure}

The velocity and temperature profiles predicted by NeuralFVM and ANSYS Fluent are presented in Fig. \ref{fig:compare_staggered_blocks}. The NeuralFVM developed in this work successfully captures the flow redistribution induced by the presence of the two internal blocks, including the velocity acceleration in the narrow passages between the blocks and the channel walls, as well as the formation of distinct wake regions downstream. The predicted high-velocity zones and low-velocity recirculation regions behind each block show good qualitative agreement with the ANSYS Fluent results. For the temperature field, NeuralFVM successfully reproduces the development of thermal boundary layers along the block surfaces and the downstream temperature distribution influenced by wake interaction and convective transport. The overall thermal patterns compare reasonably well with those obtained from ANSYS Fluent. 

\subsubsection {An array of blocks with identical size}
Upon successful verification of the preceding cases with different geometric layouts, the final demonstration case considers an array of blocks within the computational domain. Eight identical blocks are arranged in two aligned rows along the streamwise direction, with uniform spacing of $0.56m$ and between the centroid of adjacent blocks. The spanwise distance between the centroid of adjacent blocks is $0.32m$. Each block has the same geometry of $0.24m,0.16m, 0.32m$ and along the streamwise, spanwise, and vertical directions respectively. This configuration introduces repeated blockage effects along the flow direction, leading to successive wake formation and cumulative flow interactions between upstream and downstream blocks. The aligned multi-block arrangement therefore presents a more demanding test case for assessing the model \textquoteright s capability to resolve complex flow structures and transport phenomena in domains containing multiple internal solid boundaries.

\begin{figure}
    \centering
    \includegraphics[width=1.0\linewidth]{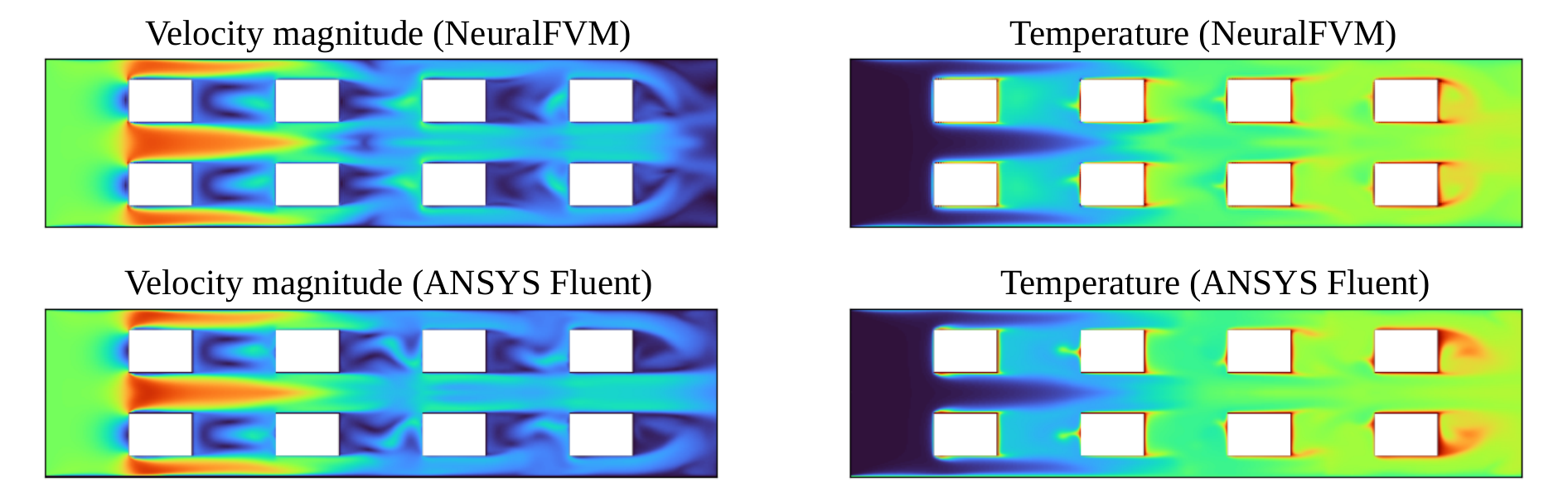}
    \caption{Comparison of velocity magnitude and temperature profiles of simulation of the array of blocks case between NeuralFVM and Fluent}
    \label{fig:compare_array_blocks}
\end{figure}

Similarly, a comparison is made between the results of NeuralFVM and ANSYS Fluent in Fig. \ref{fig:compare_array_blocks} for velocity and temperature contours at $0.08m$ above the bottom wall. The velocity contours indicate the formation of wakes behind each block and complex flow interactions as the flow moves downstream. The temperature fields exhibit similar patterns, with heat gradually spreading along the flow direction. Overall, the close agreement between the two sets of results indicates that NeuralFVM captures the main flow and thermal characteristics of the system.

\begin{figure}
    \centering
    \includegraphics[width=1.0\linewidth]{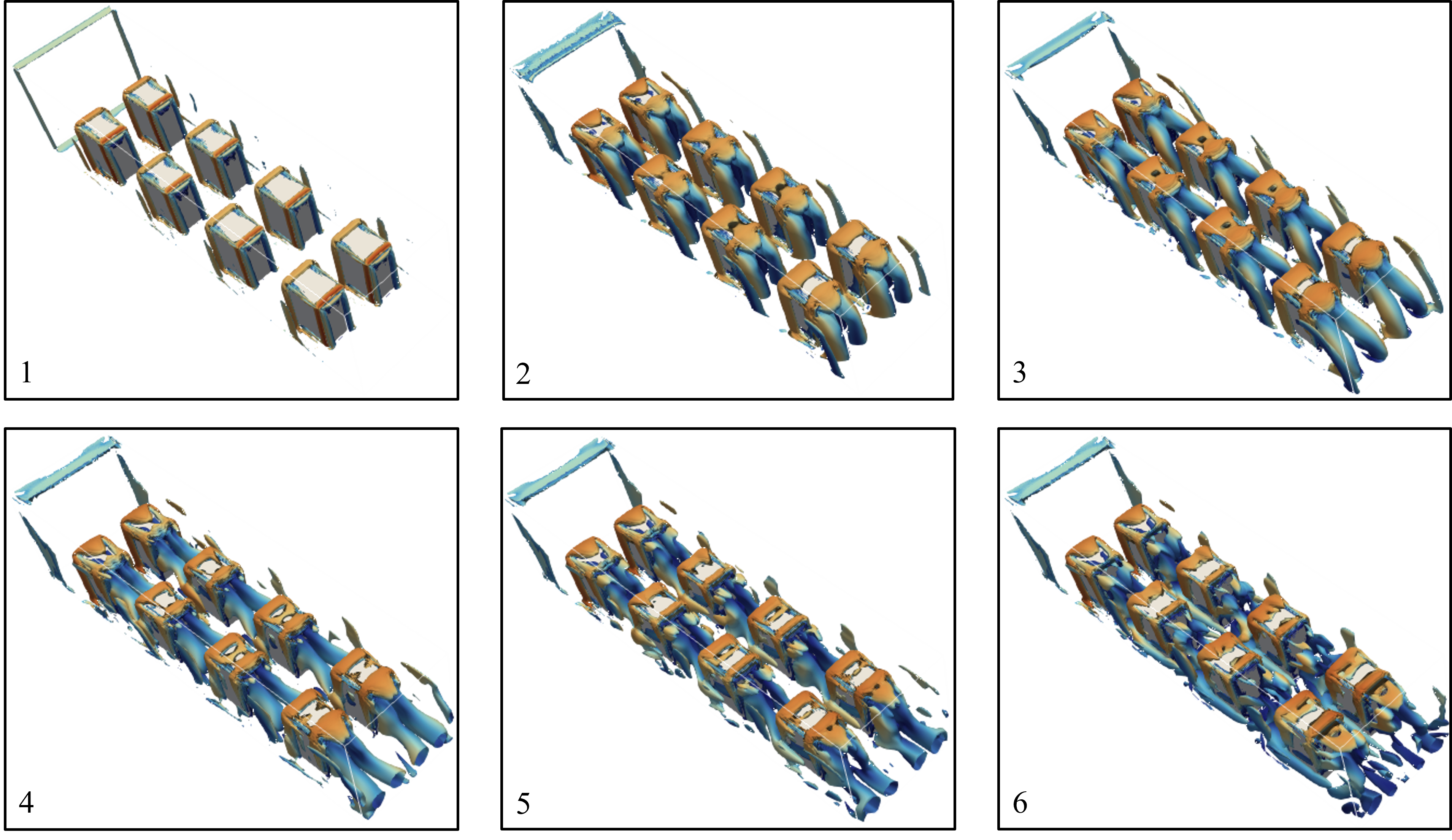}
    \caption{A sequence of plots showing the iso-contours of $\lambda_2=-0.1$ obtained from the transient NeuralFVM simulation extracted at (1) $t=0.5s$, (2) $t=3.0s$, (3) $t=5.5s$, (4) $8.0s$, (5) $10.5s$, and (6) $13.0s$}
    \label{fig:transient_lambda2}
\end{figure}

As our NeuralFVM is a transient solver, the evolution of the flow field can be easily extracted. In Fig. \ref{fig:transient_lambda2}, a sequence of $\lambda_2$ iso-contour plots is presented, where $\lambda_2$ is defined as the second largest eigenvalue of the matrix $S^2+\Omega^2$ \cite{Jeong_Hussain_1995}. A negative value indicates the presence of vortex structures. The plots show the iso-contours for $\lambda_2=-0.1$. This value used has a higher magnitude than usual so as to enhance the clarity of illustrations. It can be observed how vortical structures are generated near walls and transported downstream. 
\subsubsection{The Annex 20 benchmark indoor airflow test}
Finally, we present our simulation results for the International Energy Agency (IEA) Annex 20 indoor airflow test case \cite{annex20}, which is a benchmark problem used to validate numerical simulations of indoor airflow. Previous studies have also demonstrated that various RANS models are able to predict results that agree with the experimental results reasonably well \cite{anex20_cfx, annex20_starccm}. In this section, we present the results obtained using the NeuralFVM solver for this benchmark case based on two-dimensional (2D) simulations. We follow the common 2D setup to enable direct comparison with earlier numerical studies \cite{anex20_cfx, annex20_starccm}.  

\begin{figure}
    \centering
    \includegraphics[width=1.0\linewidth]{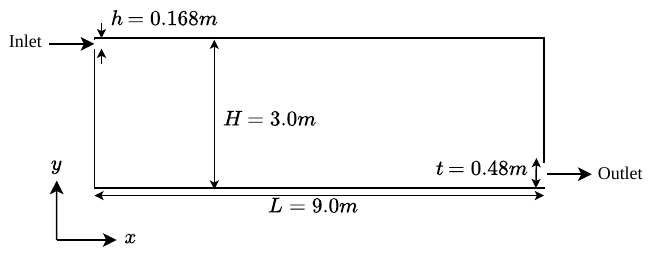}
    \caption{Sketch of 2D geometry used by the Annex 20 benchmark test}
    \label{fig:annex20_2d}
\end{figure}

The Annex 20 case considers a geometry in which air enters the computational domain through a narrow slit before developing into a room-scale airflow pattern. An illustration of the 2D geometry used is presented in Fig. \ref{fig:annex20_2d}.  It should be noted that most numerical setups consider an extension at the inlet and outlet, allowing the airflow to develop before entering the room. However, for simplicity, this extension is neglected in our setup. At the inlet, the prescribed inlet velocity is $u_0=0.455m/s$. The turbulent kinetic energy is given by $k_0=1.5(I_0u_0)^2$, where $I_0=0.04$ is the turbulent intensity. The specific dissipation is $\omega_0=\frac{\varepsilon_0}{C_\mu k_0}$, where $C_\mu=0.09$ and $\varepsilon_0=\frac{k_0^{1.5}}{l_0}$ is the dissipation rate, following \cite{annex20}, using length scale $l_0=\frac{h}{10}$. To resolve the viscous sublayer ($y^+\sim 1$), mesh size of $0.002m$ is used. The timestep used is $10^{-5}s$.

\begin{figure}
    \centering
    \includegraphics[width=1.0\linewidth]{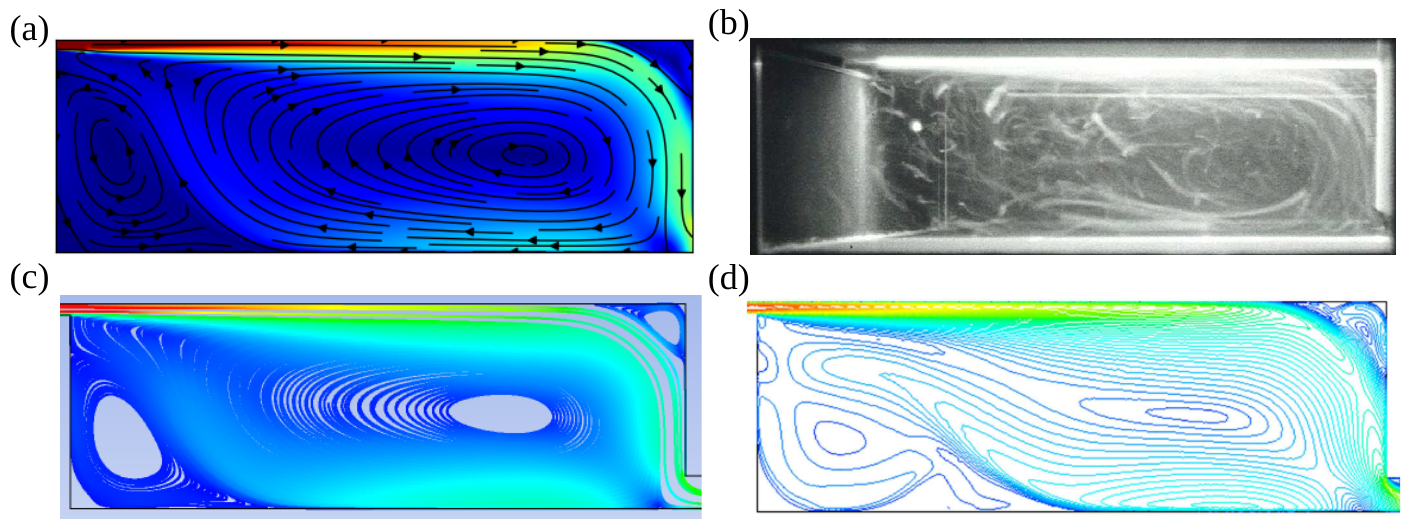}
    \caption{Results for the Annex 20 test case (a) Streamlines from the velocity field predicted by NeuralFVM (b) Experimental results from PhD thesis by Nielsen \cite{phdthesis_nielsen} (c) Streamlines from ANSYS CFX 11.0 using the standard $k$-$\omega$ model \cite{anex20_cfx} (d) Streamlines from Star-CCM+ using the standard $k$-$\omega$ model \cite{annex20_starccm} }
    \label{fig:annex20_streamlines}
\end{figure}

Figure \ref{fig:annex20_streamlines} compares the flow structures predicted by NeuralFVM with experimental observations and numerical results obtained using conventional turbulence models for the Annex 20 benchmark case. The streamlines predicted by NeuralFVM (Fig. \ref{fig:annex20_streamlines}(a)) show reasonable agreement with the experimental streaklines obtained by tracking metaldehyde particles used as flow tracers (Fig. \ref{fig:annex20_streamlines}(b)). Similar flow patterns are also obtained from simulations using the standard $k$-$\omega$ model in commercial CFD solvers (Fig. \ref{fig:annex20_streamlines}(c),(d)). In all numerical results, a large primary recirculation vortex forms near the outlet region, consistent with the experimental observations. A smaller secondary vortex develops near the bottom of the side closer to the inlet. Such small-scale vortex is not observed in experiments, possibly due to three-dimensional effects. 

\begin{figure}
    \centering
    \includegraphics[width=1.0\linewidth]{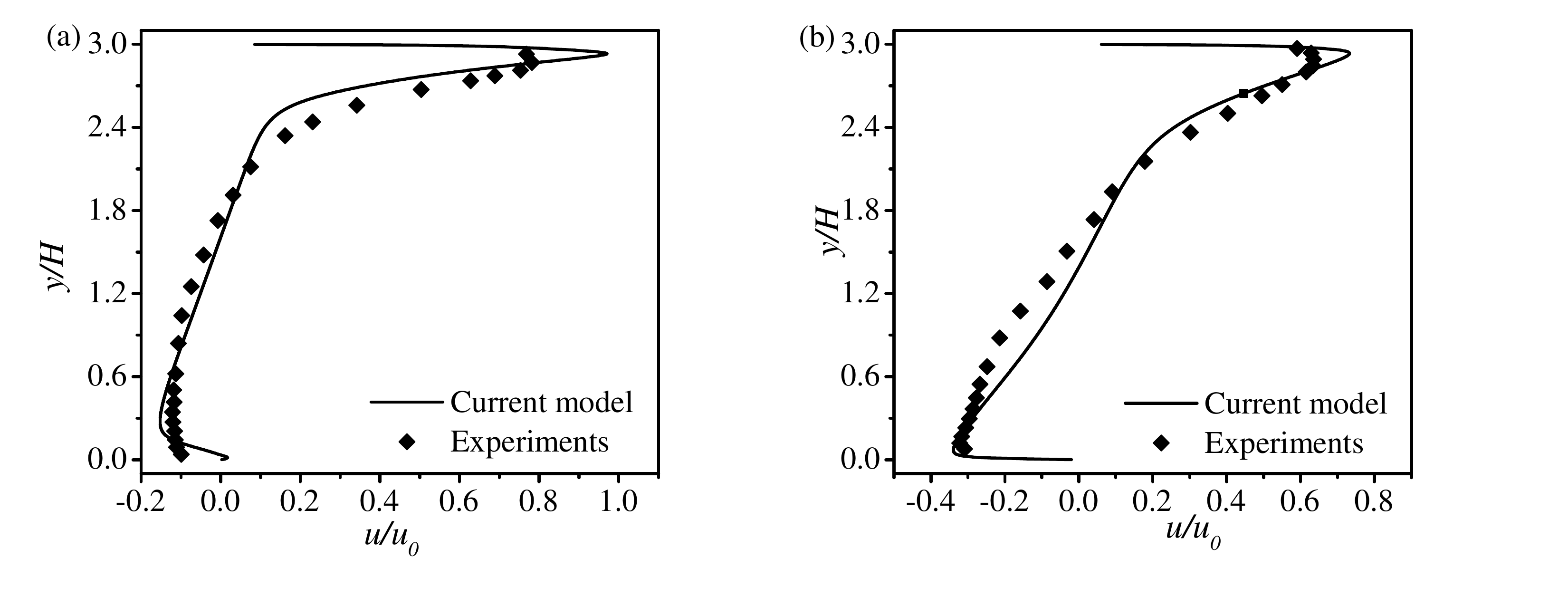}
    \caption{Comparison between the results from NeuralFVM and experimental results presented in \cite{annex20} in the symmetry plane at (a) $x=3.0m$ and (b) $x=6.0m$}
    \label{fig:annex20_quantitative}
\end{figure}

Figure \ref{fig:annex20_quantitative} compares the vertical velocity profiles with experimental measurements at two locations, i.e. $x$ = 3 m and $x$ = 6 m,  respectively. The NeuralFVM results show good agreement with the experimental data. This result further validates our GPU implementation of the standard $k$-$\omega$ turbulence model.

\subsection{Time profiling}
In this section, we demonstrate the acceleration achieved from two aspects: the use of a GPU instead of a CPU, and the use of shift operations instead of convolution operations. The total wall-clock time required to simulate 100 time steps for the open-channel verification case is measured for varying streamwise channel lengths. In all cases, the mesh size is kept constant; therefore, the total number of mesh cells increases proportionally with the channel length. These cases were performed on a computer equipped with AMD Threadripper 9970X, 32-core@4.00GHz and NVIDIA RTX PRO 5000 Blackwell, 14080 CUDA cores as the CPUs and GPUs respectively. The results are presented in Table \ref{tab:time_profile}. In the last row, we report the speedup, defined as the ratio of the CPU execution time to the GPU execution time.


\begin{table}
\centering
\caption{Time taken using CPU/GPU and using CNN/shift operation for the NeuralFVM solver to simulate 100 steps of the flow in the open-channel of different streamwise lengths.}
\label{tab:time_profile}
\begin{tabular}{|p{2.5cm}|p{1.0cm}|p{1.0cm}|p{1.0cm}|p{1.0cm}|p{1.0cm}|p{1.0cm}|}
\hline
\multirow{2}{*}{\parbox{2.3cm}{\centering Channel length\\(Mesh count)}} &
\multicolumn{2}{|p{2.0cm}|}{L=0.64m (2,097,152)} &
\multicolumn{2}{|p{2.0cm}|}{L=2.56m (8,388,608)} &
\multicolumn{2}{|p{2.0cm}|}{L=10.24m (33,554,432)} \\
\cline{2-7}
 & Shift & CNN & Shift & CNN & Shift & CNN \\
\hline
CPU & 109.4 & 191.9 & 1098.2 & 1743.3 & 3981.3 & 6011.3 \\
GPU & 5.5 & 9.8 & 22.8 & 40.0 & 140.7 & 211.2 \\
\hline
CPU/GPU & 19.9$\times$ & 19.6$\times$ & 48.2$\times$ & 43.6$\times$ & 28.3$\times$ & 28.5$\times$ \\
\hline
\end{tabular}
\end{table}

The results show that using shift operations instead of convolution operations approximately halves the simulation runtime. This improvement likely arises from  lower computational cost of tensor slicing compared with convolution operations. In addition, using a GPU instead of a CPU reduces the runtime by more than a factor of 19. It is also worth noting that NeuralFVM can be executed on either CPU or GPU architectures by modifying a single configuration variable. This flexibility represents an additional advantage of our developed solver.

\section{Conclusions}
In this work, a neural-physics-based finite volume solver, NeuralFVM, has been developed for turbulent flows using the standard $k$-$\omega$ model on GPU architectures. The governing equations are reformulated as local tensor operations, enabling seamless implementation within deep learning frameworks while preserving the conservative properties of the FVM. The proposed solver has been validated against ANSYS Fluent for a range of channel-flow configurations and indoor airflow scenarios. The results show close agreement in velocity, temperature, and turbulence quantities, demonstrating that the present approach can accurately reproduce both mean flow and turbulence characteristics. In addition, the GPU-based implementation achieves significant computational acceleration compared with its CPU counterpart, highlighting its potential for efficient large-scale simulations.

A key feature of the present framework is its fully differentiable formulation, which allows direct integration with modern ML workflows. This capability provides a promising foundation for gradient-based optimization, inverse design, and data-driven turbulence modeling. The use of local tensor operations eliminates the need for global matrix assembly, making the method well suited for highly parallel computing environments. Furthermore, the developed NeuralFVM can be executed on both GPU and CPU platforms with minimal modification to the code.

Nevertheless, the present study is conducted on uniform structured grids, which limits the range of practical applications. Further investigations involving more complex geometries, higher Re number flows, and industrial-scale applications are required to fully assess the robustness and general applicability of the proposed method. Future work will also explore the extension of the framework to more advanced turbulence models and its integration with learning-based closure strategies.

\section*{Acknowledgements}

H.Y. Li gratefully acknowledges the financial support for this project from Nanyang Technological University under the Startup Grant (SUG); the Ministry of Education (MOE), Singapore, under the MOE Tier 1 Grant (RS11/25); and the Singapore Ministry of Health through the NMRC Office, MOH Holdings Pte Ltd, under the National Epidemic Preparedness and Response R\&D Programme Funding Initiative (MOH-001041/ MOH-001073/ MOH-001446), Project No. PREPARE-OC-ETM-2025-006. The authors acknowledge the computational resources provided by the National Supercomputing Centre (NSCC) Singapore, the High Performance Computing Centre (HPCC) at Nanyang Technological University (NTU), and the School of Mechanical and Aerospace Engineering (MAE) at NTU.

\newpage
\appendix
\section{Appendix}

\subsection{Upwind transport}
The algorithm for the computing the convective flux at faces using the upwind scheme explained in Section \ref{method:fvm} is presented in Alg. \ref{algo:upwind}.
\begin{algorithm}[H]
  \caption{Upwind transport}\label{algo:upwind}
  \begin{algorithmic}[1]
    \Function{Upwind}{$\phi,u,v,w$}
        \State $\bar{u}\gets BC_x(u);\bar{v}\gets BC_y(v);\bar{w}\gets BC_z(w)$
        \State $u_{face,x}\gets L_x \ast \bar{u}$
        \State $v_{face,y}\gets L_y \ast \bar{v}$
        \State $w_{face,x}\gets L_z \ast \bar{w}$
        \State \begin{varwidth}[t]{\linewidth} $(u\phi)^{upwind}_{face,x}\gets \left((W_x^+\ast BC_x(\phi)) \odot I[u_{face,x}\geq 0] \right.$\par 
        \hskip\algorithmicindent $\left.+(W_x^-\ast BC_x(\phi))\odot I[u_{face,x} < 0]\right) \odot u_{face,x}$
        \end{varwidth}
        \State \State \begin{varwidth}[t]{\linewidth} $(v\phi)^{upwind}_{face,x}\gets \left((W_y^+\ast BC_y(\phi)) \odot I[v_{face,y}\geq 0] \right.$\par 
        \hskip\algorithmicindent $\left.+(W_y^-\ast BC_y(\phi))\odot I[v_{face,y} < 0]\right) \odot v_{face,y}$
        \end{varwidth}
        \State \State \begin{varwidth}[t]{\linewidth} $(w\phi)^{upwind}_{face,z}\gets \left((W_z^+\ast BC_z(\phi)) \odot I[w_{face,z}\geq 0] \right.$\par 
        \hskip\algorithmicindent $\left.+(W_z^-\ast BC_z(\phi))\odot I[w_{face,z} < 0]\right) \odot w_{face,z}$
        \end{varwidth}
       \State \Return $(u\phi)^{upwind}_{face,x},(v\phi)^{upwind}_{face,y},(w\phi)^{upwind}_{face,z}$
    \EndFunction
  \end{algorithmic}
\end{algorithm}

\newpage
\subsection{Navier-Stokes time stepping}\label{appendix:ns_timestep}
The FVM is applied on the simple example of the laminar Navier-Stokes equations in Alg. \ref{algo:nse_explicit}. The steps can generally be broken down into computing the face-centered values, then integrating them using the divergence theorem.
\begin{algorithm}[H]
  \caption{Explicit component of Navier-Stokes equations}\label{algo:nse_explicit}
  \begin{algorithmic}[1]
    \Function{$F^{NSE}$}{$u,v,w,p,T$}
         \State $\bar{u}\gets BC_x(u);\bar{v}\gets BC_y(v);\bar{w}\gets BC_z(w)$
         \State $p\vert_{face,x}\gets L_x\ast BC_x({p})$
         \State $p\vert_{face,y}\gets L_y\ast BC_y(p)$
         \State $p\vert_{face,z}\gets L_z\ast BC_z(p)$
        \For{$\phi \in \{u,v,w\}$}
        \State $(u\phi)^{upwind}_{face,x},(v\phi)^{upwind}_{face,y},(w\phi)^{upwind}_{face,z} \gets \Call{Upwind}{u,u,v,w}$
        \State $\frac{\partial \phi}{\partial x}\vert_{face,x}\gets D_x\ast BC_x(\phi)$
        \State $\frac{\partial \phi}{\partial y}|_{face,y}\gets D_y\ast BC_y(\phi)$
        \State $\frac{\partial \phi}{\partial z}\vert_{face,z}\gets D_z\ast BC_z(\phi)$
        \EndFor
        \For{$\phi \in \{u,v,w\}$}
      \State \begin{varwidth}[t]{\linewidth} $\frac{\partial \phi}{\partial t}\gets D_x\ast\left(-(u\phi)_{face,x}^{upwind}+\nu\frac{\partial \phi}{\partial x}|_{face,x}-p\vert_{face,x}I[\phi=u]\right)$\par
        \hskip\algorithmicindent $+D_y\ast \left(-(v\phi)_{face,y}^{upwind}+\nu \frac{\partial \phi}{\partial y}|_{face,y}-p\vert_{face,y}I[\phi=v]\right)$\par
        \hskip\algorithmicindent$+D_z\ast\left(-(w\phi)_{face,z}^{upwind}+\nu\frac{\partial \phi}{\partial z}|_{face,z}-p\vert_{face,z}I[\phi=w]\right)$
      \end{varwidth}
\EndFor
    \State \Return $\frac{\partial u}{\partial t}, \frac{\partial v}{\partial t}, \frac{\partial w}{\partial t}$
    \EndFunction
  \end{algorithmic}
\end{algorithm}
\newpage
\subsection{Semi-implicit destruction of $k$-$\omega$}
The algorithm for advancing the destruction term of the $k$ adnd $\omega$ transport equations are presented in Alg. \ref{algo:komega_implicit}. 

\begin{algorithm}[H]
  \caption{Semi-implicit component of $k$-$\omega$}\label{algo:komega_implicit}
  \begin{algorithmic}[1]
    \Function{$F^{k\text{-}\omega}_{implicit}$}{$u,v,w,k,\omega,\Delta t$}
         \For{$\phi \in \{u,v,w,k,\omega\}$}
            \State $\frac{\partial \phi}{\partial x}\vert_{face,x}\gets D_x\ast BC_x(\phi)$
            \State $ \frac{\partial \phi}{\partial y}|_{face,y}\gets D_y\ast BC_y (\phi)$
            \State $\frac{\partial \phi}{\partial z}\vert _{face,z}\gets D_z\ast BC_z(\phi)$
         \State $\frac{\partial \phi}{\partial x}\gets L_x\ast \frac{\partial \phi}{\partial x}\vert_{face,x}$
         \State $\frac{\partial \phi}{\partial y}\gets L_y\ast \frac{\partial \phi}{\partial y}\vert_{face,y}$
         \State $\frac{\partial \phi}{\partial z}\gets L_z \ast \frac{\partial \phi}{\partial z}\vert_{face,z}$
        \EndFor
         \State $S_{ij}\gets \frac{1}{2}(\frac{\partial u_i}{\partial x_j}+\frac{\partial u_j}{\partial x_i})$ 
         \State $\Omega_{ij}\gets \frac{1}{2}(\frac{\partial u_i}{\partial x_j}-\frac{\partial u_j}{\partial x_i})$ 
          \State $\chi_k\gets \frac{1}{\max(\omega,\varepsilon)^3}\left(\frac{\partial k}{\partial x}\frac{\partial \omega }{\partial x}+\frac{\partial k}{\partial y}\frac{\partial \omega }{\partial y}+\frac{\partial k}{\partial z}\frac{\partial \omega }{\partial z}\right)$
          \State $f_{\beta^*}\gets \begin{cases}
              1.0,& \chi_k \leq 0\\
              \frac{1+680\chi_k^2}{1+400\chi_k^2},& \chi_k>0
          \end{cases}$
          \State $Re_t \gets \frac{k}{\nu \max(\omega,\varepsilon)}$
          \State $\beta^* \gets \beta_{\infty}^* \frac{\frac{4}{15}+(\frac{Re_t}{R_\beta})^4}{1+(\frac{Re_t}{R_\beta})^4}$
          \State $k^{new}\gets \frac{k}{1+\Delta t \beta^* f_{\beta^*}\omega}$
          \State $\chi_\omega \gets |\frac{\Omega_{ij}\Omega_{jk}S_{ki}}{(\beta_\infty^* \max(\omega,\varepsilon)^3)}|$
          \State $f_\beta \gets \frac{1+70\chi_{\omega}}{1+80\chi_\omega}$
          \State $\omega^{new} \gets \frac{\omega}{1+\Delta t \beta f_{\beta}\omega}$
          \State \Return $k^{new},\omega^{new}$
    \EndFunction
  \end{algorithmic}
\end{algorithm}

\subsection{Explicit time stepping of the rest of the terms}
The advancement of the rest of the terms other than the destruction terms in the $k$-$\omega$ governing equation can be done in a way similar to \ref{appendix:ns_timestep} by computing the relevant terms in a sequential manner. Measures are taken to ensure stability of the terms, such as clamping the values of $k$ and $\omega$ so that they remain positive.

\FloatBarrier
 \bibliographystyle{elsarticle-num} 
 \bibliography{ref.bib}
\end{document}